\documentclass[11pt,reqno]{amsproc}
\usepackage[colorinlistoftodos,shadow]{todonotes}
\usepackage[T1]{fontenc}
\usepackage{amsmath,amscd,amssymb}
\usepackage{cite}
\usepackage{color}
\usepackage{graphicx}
\usepackage{psfrag,epsfig}
\usepackage{multirow}
\usepackage{rotating}
\usepackage{fullpage}
\usepackage{subfigure}
\usepackage{enumerate}
\usepackage{comment}
\usepackage{lineno}
\usepackage{algorithmic}
\usepackage{algorithm}
\usepackage[small]{caption}
\usepackage[debug=false, colorlinks=true, pdfstartview=FitV, 
linkcolor=blue, citecolor=blue, urlcolor=blue]{hyperref}
\newtheorem{theorem}{Theorem}[section]

\newtheorem{remark}[theorem]{Remark}

\numberwithin{equation}{section}
\usepackage{chemarr}
\usepackage[version=3]{mhchem}
\newcommand*\patchAmsMathEnvironmentForLineno[1]{%
  \expandafter\let\csname old#1\expandafter\endcsname\csname #1\endcsname
  \expandafter\let\csname oldend#1\expandafter\endcsname\csname end#1\endcsname
  \renewenvironment{#1}%
     {\linenomath\csname old#1\endcsname}%
     {\csname oldend#1\endcsname\endlinenomath}}%
\newcommand*\patchBothAmsMathEnvironmentsForLineno[1]{%
  \patchAmsMathEnvironmentForLineno{#1}%
  \patchAmsMathEnvironmentForLineno{#1*}}%
\AtBeginDocument{%
\patchBothAmsMathEnvironmentsForLineno{equation}%
\patchBothAmsMathEnvironmentsForLineno{align}%
\patchBothAmsMathEnvironmentsForLineno{flalign}%
\patchBothAmsMathEnvironmentsForLineno{alignat}%
\patchBothAmsMathEnvironmentsForLineno{gather}%
\patchBothAmsMathEnvironmentsForLineno{multline}%
}

\title{Sequential geophysical and flow inversion to characterize fracture
networks in subsurface systems}
\author{M.~K.~Mudunuru$^{*}$, S.~Karra, N.~Makedonska, and T.~Chen \\
{\scriptsize Earth and Environmental Sciences Division, 
Los Alamos National Laboratory, Los Alamos, NM 87545.} \\
}
\address{\small $^*$Corresponding author: Dr.~Maruti Kumar Mudunuru, 
Computational Earth Science Group (EES-16), Earth and Environmental Sciences 
Division, Los Alamos National Laboratory, Los Alamos, NM 87545 \\
\textbf{E-mail address:} maruti@lanl.gov \newline \newline
}
\date{\today}
\begin{document}
\maketitle
%
%
\section*{Abstract} 
Subsurface applications including geothermal, geological carbon 
sequestration, oil and gas, etc., typically involve maximizing 
either the extraction of energy or the storage of fluids.
Fractures form the main pathways for flow in these systems
and locating these fractures is critical for predicting flow.
However, fracture characterization is a highly uncertain process
and data from multiple sources, such as flow and geophysical, 
is needed to reduce this uncertainty.
We present a non-intrusive sequential inversion framework, 
for integrating data from geophysical and flow sources to 
constrain fracture networks in the subsurface. In 
this framework, we first estimate bounds on the statistics 
for the fracture orientations using microseismic data. 
These bounds are estimated through a combination of a focal 
mechanism (physics-based approach) and clustering analysis 
(statistical approach) of seismic data. Then, the fracture 
lengths are constrained using flow data. The efficacy 
of this inversion is demonstrated through a representative example.
\\
\\
\textbf{Keywords:} sequential inversion, multiple datastreams, 
geophysics, flow, fracture, subsurface modeling, clustering analysis, $k$-means
clustering, Latin hypercube sampling, elbow method.
%

\section{INTRODUCTION}
\label{Sec:S1_JI_Intro}
The efficiency of subsurface applications such as 
unconventional oil and gas, $\mathrm{CO}_2$ sequestration, 
waste water disposal, and geothermal systems \cite{1996_NRC_Fractures,
2016_Mudunuru_SGW,2017_Mudunuru_etal_arXiv},
where fluids are injected or extracted, vastly depends on the 
permeability of the subsurface. Fractures in the subsurface form
critical flow pathways and significantly influence the permeability.
Due to the complex (heterogeneous and anisotropic) nature of the subsurface, 
there is a lot of uncertainty involved in the characterization of these
fracture networks \cite{1996_NRC_Fractures,
2004_Cosgrove_Engelder,Brown_etal}. Constraining subsurface fracture
networks and their interaction with fluid flow is one of the great 
challenges in the earth and energy 
sciences \cite{2005_Shaw,2007_Eberhardt_etal1,2007_Eberhardt_etal2,
2014_Spence_etal}. 

The uncertainty in characterizing fracture networks is severe due to the 
following reasons: 1) typically, fractures are located at around 10000 
feet below the surface of earth and are not easily accessible;
2) the geometry and topology of these networks, 
which control various processes in the subsurface 
system such as flow, stress, heat, reactions, etc., are unknown; 3) 
existing methods use a single data source (for instance, either flow
or geophysical data) for characterization. Using multiple datasets
helps in reducing the uncertainty and provides better constraints
on the fracture topology and geometry. However, the main 
challenge with using multiple datasets is that 
a typical time-series subsurface data set is rough and highly-oscillatory, 
with missing details \cite{Brown_etal,2014_Zhou_etal_AIP,HDR_FentonHill_2015}. 
Furthermore, there is a strong correlation between various data 
streams such as seismic, flow, temperature, 
geochemical and the subsurface fracture 
network parameters \cite{1984_Robinson_Tester_JGR_v89_p10374_p10384,
1989_Grisby_etal_Geothermics_v18_p629_p656,1989_Grisby_etal_Geothermics_v18_p657_p676,
1993_Sahimi_etal_PRL_p2186_p2189,1996_Roff_etal_IJRMMSGA_v25_p627_p639,
2014_Pandey_etal_Geothermics_v51_p46_p62,2015_Pandey_etal_Geothermics_v51_p46_p62,
2014_McClure_Horne_IJRMMS_v72_p242_p260,2016_Kang_etal_WRR_v52_p903_p919,
2006_Chen_WRR_v42,2012_Dorn_WRR_v48}, and 
thus these data streams must be considered to accurately parameterize the
fracture networks. 
Previous methods, which are primarily based on outcrop analysis, 
had difficulty in accounting for these correlations \cite{2011_Geier_Forsmark,2011_SKB,
2013_Hartley_Joyce_JH_v500_p200_p216}. Our inversion method uses
recently developed clustering analysis algorithms to overcome this
barrier. Our aim, in this paper, 
is to develop a non-intrusive sequential
inversion framework, using 
multiple types of data (specifically, geophysical and flow 
datasets) to constrain subsurface 
fracture networks.
The output of the inversion is a discrete fracture network (DFN) that 
models the fracture system as a network of two-dimensional
planes in three-dimensional space. 
The next subsection details the assumptions made and the
observational data needed in our framework.

\subsection{Subsurface fracture/fault statistics:~Assumptions, basic 
  workflow, and constraints}
\label{SubSec:S1_JI_FS}
Microseismicity occurrence caused due to fluid 
injection may be correlated with the subsurface  
fracture networks \cite{1996_Roff_etal_IJRMMSGA_v25_p627_p639,
2015_Shapiro}. Microseismic events are typically 
recorded during the various stages of 
field-scale stimulation in reservoirs \cite{2012_NAP} and cluster of these microseismic 
events can be used to identify a connected subsurface fracture 
network \cite{1999_Talebi,1999_Wyss_etal}. Here, we briefly 
describe how the microseismic data combined with flow data, 
can be used to characterize the fracture network. 

Based on the recording of the microseismic events, velocity models 
are constructed.  These velocity models are key components \cite{2011_Refunjol_Marfurt_LE,
2013_Michelena_etal_LE,2016_Dadi_etal} for locating earthquake events/hypocenters. 
Various methods exist in literature to construct seismic velocity 
models \cite{2008_IanJones_etal}. Within the oil and gas 
industry, full waveform inversion (FWI) is now the state-of-art 
seismic velocity building algorithm \cite{1984_Tarantola_Geo_v49_p1259_p1266,
2009_Virieux_Operto_Geo_v74}. FWI is a nonlinear optimization problem 
which iteratively updates the velocity model to reduce misfit between 
the recorded and synthesized seismic data via the adjoint method. Once 
a velocity model is constructed, we invert for microseismic event 
locations using seismic wave arrival times. 

The accuracy of locating microseismic events is controlled by various 
factors such as fracture network, fracture geometry, knowledge of the earth's 
crust, and the measurement accuracy of seismic wave arrival-times. The earthquake 
location methods generally incorporate absolute travel-time measurements
and/or cross-correlation P-wave and S-wave differential travel-time 
measurements. Residuals between observed and theoretical travel-time 
differences between a pair of events are minimized. This approach of 
estimating earthquake hypocenters is called double-difference earthquake 
location algorithm \cite{2000_Waldhauser_etal_BSSA_v90_p1353_p1368}. This algorithm is 
used our paper to invert for seismic event locations. Once the event 
locations are inverted, we can use seismic waveform information to 
obtain focal mechanisms. These focal mechanisms provide information 
on fracture orientation. See Section \ref{Sec:S2_JI_GE} for more details.

We then assume the lower and upper bounds on the fracture length in this 
paper. However, fracture density shear-wave splitting analysis can be used
to provide these bounds through fracture compliance and fracture length relationship 
\cite{2013_Verdon_Wustefeld_GP_v61_p461_p475}. This splitting 
method is based on the fact that a shear-wave propagating 
through an isotropic elastic solid containing stress-aligned 
micro-cracks behaves as if the solid was anisotropic 
\cite{2003_Vlahovic_Elkibbi_Rial_JVGR_v120_p123_p140}. 
The shear-wave splits into two waves, a fast one polarized 
parallel to the predominant crack direction and a slow one 
polarized perpendicular to it. The time delay between the 
arrivals of the fast and the slow waves is proportional 
to crack density, or number of cracks per unit volume 
along the wave path.

Fracture pattern, connectivity, and size distribution can be obtained 
from fractal analysis\footnote{This approach gives relationships between 
number of fractures $N_l$ and length scale $l$ for fracture surfaces, 
which is given by: $N_l \propto l^{-D}$, where $D$ is the fractal 
dimension. $D \approx 1.9$ and 2.5, in two and three dimensions, 
respectively. The fractal dimension $D$ is an indicator for material 
heterogeneity and strength of earthquakes. If the total number of 
fractures are known, then an estimate on the average length of the 
fractures can be obtained.}, $b$-value analysis\footnote{$b$-value 
is a statistic measuring the proportions of large and small earthquakes in a seismic 
event cluster. If $b$-value is large, small earthquakes are relatively common, whereas 
when $b$-value is small, there exist a possibility of a large event 
happening \cite{2015_Srivastava_etal_NH_v77_p97_p107}. It should be 
noted that there is a strong correlation between $b$-value and the 
fractal dimension $D$.}, and clustering analysis\footnote{Cluster 
analysis is the task of grouping a set of objects in such a way that objects 
in the same group are more similar to each other than to those in other groups. 
Through clustering analysis of seismic data, one can obtain attributes such as 
mean orientation of a fracture/fault. Furthermore, clustering of microseismic data 
also helps to find the most likely and significant locations of the fractures 
\cite{1993_Sahimi_etal_PRL_p2186_p2189}.} of microseismic events 
\cite{1994_Sahimi,2011_Sahimi,1993_Sahimi_etal_PRL_p2186_p2189}. We use
clustering analysis, specifically $k$-means clustering, in this paper. 
In addition to microseismic data, 
orthogonal information such as flow rate, pressure, tracer, temperature, 
and geochemical datasets can then be used to constrain the fracture length/size.
In this paper, we use pressure data, for constraining fracture lengths.
Once we get the fracture statistics, 
permeability can be estimated, that then can be utilized to obtain the 
state of stress in a reservoir. State of stress can then be 
used to gain insight for drilling new wells and in assessing risk for decision-making.

\subsubsection{\textbf{Assumptions behind the proposed inversion framework}}
The following are the main assumptions behind 
our sequential geophysical and flow inversion framework to estimate 
fracture statistics such as fracture orientation and fracture 
size distributions:
\begin{itemize}
  \item [$\blacktriangleright$] The fluid flow in a fault damage 
    zone is assumed to be predominantly within its (background 
    and fault-related) fracture networks, which mostly consists 
    of several fault-related fracture sets \cite{Brown_etal,
    2010_Manzocchi_etal_Geofluids_v10_p94_p113,
    2002_Berkowitz_AWR_v25_p861_p884,2016_Morley_Nixon_JSC_v84_p68_p84,
    2000_Aydin_MPG_v17_p797_p814,1993_Moreno_Neretnieks_JCH_v14_p163_p192,
    2015_Kang_etal_WRR_v51_p940_p959}. This means, the virgin rock (also called 
    as the matrix) is assumed 
    to be impermeable. Relaxing this assumption and accounting for 
    fluid loss due to diffusion/transport of fluid in and out of the matrix is 
    beyond the scope of the current paper.
  \item [$\blacktriangleright$] Clustering of microseismic 
    events represents a connected fault-fracture network 
    \cite{2006_Hofrichter_Winkler_EG_v49_p821_p827}. Each cluster of 
    microseismic events may correspond to a unique fracture or a set 
    of fractures which have similar spatial and hydrological attributes. 
    This means that, across various fractures within each cluster, the 
    statistical variance in material and geometrical properties such 
    as fracture permeability, aperture, and transmissivity is not high.
  \item [$\blacktriangleright$] Cluster centers of microseismic 
    events are assumed to be the observation points where the 
    flow and pressure data are sampled and/or extracted and/or 
    monitored over time. For instance in case of enhanced geothermal 
    systems, cluster centers may represent the possible location 
    of new drilling (injection and/or production) wells \cite{Brown_etal,
    2014_NICTA,1996_Roff_etal_IJRMMSGA_v25_p627_p639,
    2012_Ghassemi_GGE_v30_p647_p664,2013_Ziagos_etal_SGW,
    2014_McClure_Horne_IJRMMS_v72_p242_p260}. It should be 
    noted that the cluster centers may move in time, representing 
    fracture propagation or fluid movement within the subsurface 
    fracture network \cite[Chapter-4]{1994_Sahimi} 
    \cite[Chapter-6 and Chapter-7]{2011_Sahimi}. Such an analysis 
    where we account for time-series cluster centers based on the 
    state of earth's stress is beyond the scope of current paper. 
\end{itemize}

\subsection{Study objective and outline of the paper}
\label{SubSec:S1_JI_Scope}
The main objective of this study is to develop a sequential inversion 
framework to constrain subsurface fracture network (or equivalent DFN) stochastics. 
The paper is organized as follows: Section \ref{Sec:S2_JI_GE} describes the seismic wave propagation 
and flow forward models used in the sequential inversion framework. From 
this inversion process, we constrain the major fracture orientation 
and fracture lengths of subsurface DFNs. 
Assumptions in modeling these systems are also outlined. In Section 
\ref{Sec:S3_JI_Results}, we present a synthetic example to illustrate 
our sequential inversion framework to constrain the fracture statistics, 
which is orientation and lengths of DFNs. We also show how clustering 
analysis of seismic events can augment the seismic inversion methods in 
better constraining the fracture orientation. Pressure data sets
at certain observation points (which can be injection, production or 
observation wells 
in real life applications) are used to constrain the fracture length in the 
synthetic example. Finally, conclusions are drawn in Section 
\ref{Sec:S4_JI_Conclusions}.

%

\section{FORWARD MODEL AND SEQUENTIAL INVERSION FRAMEWORK}
\label{Sec:S2_JI_GE}
In this section, we briefly describe the forward models for constructing 
focal mechanisms as well as the governing equations for fluid flow. We then present 
a numerical methodology to construct our sequential geophysical 
and flow inversion framework. The framework is constructed based on 1) 
\textsf{dfnWorks} \cite{2014_Hyman_etal_SIAMJSC_v36_pA1871_pA1894,
2015_Hyman_etal_CG_v84_p10_p19,2015_Makedonska_etal_CG_v19_p1123_p1137, 
2015_Karra_etal_WRR_v51_p8646_p8657,2015_Hyman_etal_WRR_v51_p7289_p7308}, 
which is a parallelized computational suite to generate three-dimensional 
DFNs and simulate flow and transport 
on these networks; 2) 
\textsf{MADS} software \cite{2016_MADS} for constructing Latin hypercube 
samples (LHS) of various fracture network parameters. \textsf{MADS} is an open-source 
high performance computational framework for data-\& model-based analyses.

\subsection{Forward model:~Focal mechanisms}
\label{SubSec:S2_JI_FM_FocMech}
Focal mechanisms of microseismic events describe the seismic 
source motions on the fault-related fractures, and can provide 
useful information on their orientations. Given adequate seismic 
records, knowledge of velocity models and event locations, we 
can invert the focal mechanism for each microseismic event to  
constrain fracture orientation. This is achieved by constructing
true seismic waveforms and arrival times of both compressional 
and shear waves for each event based on event location, focal 
mechanisms, and velocity model built upon seismic wave propagation 
theory \cite{2002_Aki_Richards,2004_Chapman,2009_Shearer,2012_Sato_etal}. 
The inverted focal mechanism parameters, strike angle and dip angle, 
are directly related to fracture orientation. Strike angle describes 
the direction of a fracture relative to North in the clockwise direction 
and dip angle describes the direction of a fracture relative to horizon 
in the clockwise direction.

In this paper, assuming a velocity model, we first invert for microseismic 
event locations using seismic wave arrival times. A double-difference 
event location algorithm \cite{2000_Waldhauser_etal_BSSA_v90_p1353_p1368} 
is used to invert the event locations. From the obtained inverted 
event locations, we can use seismic waveform information to obtain 
focal mechanisms. Our waveform focal mechanism inversion method 
\cite{2014_Chen_etal_JGRSE_v119_p3035_p3049} inverts for 7 parameters 
for each event:~Strike angle, dip angle, slip, isotropic, Compensated 
Linear Vector Dipole (CLVD) component, source duration, and seismic 
moment. Green's functions are calculated numerically to simulate seismic 
waveforms for given event location, focal mechanism, and velocity model. 
Inverted event focal mechanism is obtained by minimizing the misfit between 
true and simulated seismic waveforms using a simulated heal annealing method. 

\subsection{Forward model:~Fluid flow}
\label{SubSec:S2_JI_FM_FocMech}
The forward model for the fluid flow is based on single phase, 
fully saturated, and isothermal Richards equation \cite[Chapter-11]{Pinder_Celia}. 
The governing mass conservation equation for fully saturated 
fluid flow is given as follows:
\begin{align}
  \label{Eqn:VarSat_Richards_Eqn}
  \frac{\partial \left(\varphi \rho \right)}{\partial t} + 
  \mathrm{div}[\rho \mathbf{v}] = Q_w
\end{align}
where $t$ denotes the time, $\varphi$ denotes the porosity of 
the porous medium, $\rho$ is the fluid density $[\mathrm{kg} \, 
\mathrm{m}^{-3}]$, $\mathbf{v}$ is the Darcy's velocity $[\mathrm{m} 
\, \mathrm{s}^{-1}]$, and $Q_w$ is the volumetric source/sink term 
$[\mathrm{kg} \, \mathrm{m}^{-3} \, \mathrm{s}^{-1}]$. The Darcy's 
velocity is given as follows:
\begin{align}
  \label{Eqn:Darcy_Flux_Eqn}
  \mathbf{v} = - \frac{k_s}{\mu} \mathrm{grad}
  [p - \rho g z]
\end{align}
where $k_s$ is the saturated permeability $[\mathrm{m}^{2}]$, $\mu$ 
is the dynamic viscosity $[\mathrm{Pa} 
\, \mathrm{s}]$, $p$ is the fluid pressure $[\mathrm{Pa}]$, $g$ is 
the gravity $[\mathrm{m} \, \mathrm{s}^{-2}]$, and $z$ is the vertical 
component of the position vector $[\mathrm{m}]$. 

The nonlinear partial differential equations \eqref{Eqn:VarSat_Richards_Eqn}--\eqref{Eqn:Darcy_Flux_Eqn} 
describing the  fluid flow on discrete fracture networks are solved 
using the massively parallel subsurface flow simulator \textsf{PFLOTRAN} \cite{2015_Lichtner_etal_PFLOTRAN}, 
which employs a fully implicit backward Euler for discretizing time 
and a two-point flux finite volume method for spatial discretization. The resulting non-linear 
algebraic equations are solved using a Newton-Krylov solver. 

\subsection{Sequential inversion framework}
\label{SubSec:S2_JI_Framework}
In order to constrain the fracture orientation, we first construct 
focal mechanisms of microseismic events. Once the location of these 
events are obtained, we perform cluster analysis to obtain the corresponding 
discrete probability distributions for strike and dip angles. In 
addition, we also obtain the cluster centers (which are the observation 
points \cite{2015_Okada_etal_Geofluids_v15_p293_p309}) where the data 
for the flow is sampled. More details on the numerical methodology to 
constrain fracture orientation are provided in Algorithm \ref{Algo:DFN_Orientation}. 

\begin{algorithm} 
  \caption{{\small A numerical methodology to constrain fracture orientation 
    for discrete fracture networks}}
  \label{Algo:DFN_Orientation} 
  \begin{algorithmic}[1]
    \STATE INPUT:~Focal mechanisms; locations of microseismic events 
      (\texttt{Coord}); total number of events (\texttt{NumCoord});
      maximum number of iterations to run $k$-means clustering algorithm 
      to return a codebook of microseismic events with lowest distortion
      (\texttt{MaxIters}); tolerance/threshold after which $k$-means algorithm 
      is terminated if the change in distortion from the last $k$-means iteration 
      is less than or equal to (\texttt{TolValue}); and maximum combinations 
      allowed for three-point computational geology problem (\texttt{MaxCombo}) 
      \cite{2000_Vacher_JGE_v48_p522_p531}.
      \begin{itemize}
         \item \texttt{Coord} contains $(x,y,z)$ coordinates of microseismic events.
      \end{itemize}
    \IF {$\frac{\mathrm{\texttt{NumCoord}}!}{\left(\mathrm{\texttt{NumCoord}} 
      - 3 \right)! \, 3!} \; \; < \; \; \mathrm{\texttt{MaxCombo}}$}
      \STATE Construct strike and dip angles discrete probability distributions 
        for entire \texttt{Coord}.
      \STATE For each combination of three points in \texttt{Coord}, 
        construct the $x$, $y$, and $z$ coefficients of the fracture 
        plane \cite{2000_Vacher_JGE_v48_p522_p531}. Let the coefficients 
        be denoted as $\mathfrak{a}$, $\mathfrak{b}$, and $\mathfrak{c}$.
      \STATE Determine the strike and dip angles for trivial cases (that is, 
        when one or more coefficients of the fracture plane are equal to zero).
      \STATE For non-trivial cases, we have the following:
        \IF {$\mathfrak{a} \neq 0$}
          \STATE Strike angle = $\frac{180}{\pi} \arctan \left(-
            \frac{\mathfrak{b}}{\mathfrak{a}} \right)$.
        \ENDIF
        \IF {Strike angle < 0}
          \STATE Strike angle = $180^{\mathrm{o}}$ - absolute value of previously 
            obtained strike angle (In this case, note that we have counter clockwise 
            strike angle orientation).
        \ENDIF
        \IF {$\mathrm{Sign}[- \mathfrak{c} \times \mathfrak{a}] > 0$}
          \STATE Dip angle = $180^{\mathrm{o}}$ - $\frac{180}{\pi} 
            \arctan \left(\sqrt{\frac{\mathfrak{a}^2 + \mathfrak{b}^2}{
            \mathfrak{c}^2}} \right)$.
        \ELSE
          \STATE Dip angle = $\frac{180}{\pi} \arctan \left(\sqrt{
            \frac{\mathfrak{a}^2 + \mathfrak{b}^2}{\mathfrak{c}^2}} 
            \right)$.
        \ENDIF 
    \ENDIF
    \STATE Determine the number of clusters `$k$' for $k$-means algorithm using 
      a combination of elbow method \cite{2011_Everitt_etal}, focal mechanisms, 
      and from (strike and dip angles) discrete probability distributions obtained 
      from the entire \texttt{Coord} (if computationally tractable).
    \STATE Cluster the microseismic data and generate a codebook for these
      events using $k$-means clustering and vector quantization algorithms
      (herein, implementation is based on \textsf{Scipy.Cluster} module 
      \cite{2016_Scipy} and parallelization is performed using \textsf{Multiprocessing 
      Python} module \cite{2016_MP_Python}). Inputs for these algorithms are 
      \texttt{Coord}, \texttt{MaxIters}, and \texttt{TolValue}.
    \FOR{$i = 1, 2, \cdots, k$}
      \STATE Construct strike and dip angles discrete probability distributions 
      for each set of $\mathrm{\texttt{Coord}}_i$, where $\mathrm{\texttt{Coord}}
      _i$ are the coordinates of microseismic events of cluster $i$.
      \STATE Analysis is similar to that of the full cluster coordinates $\mathrm{\texttt{Coord}}$, which 
        is discussed in an earlier if-else statement
    \ENDFOR
    \STATE OUTPUT:~$\mathrm{\texttt{Coord}}_i$, $\mathrm{\texttt{ClusterCentroid}}_i$, 
      strike angles, and dip angles discrete probability distributions for each cluster, 
      where $i = 1, 2, \cdots, k$.
  \end{algorithmic}
\end{algorithm}

\begin{algorithm} 
  \caption{{\small A numerical methodology to constrain fracture 
    length for discrete fracture networks}}
  \label{Algo:DFN_Length} 
  \begin{algorithmic}[1]
    \STATE INPUT: Number of fractures ($n$); domain size; 
      centroids of the 
      user-defined fracture ellipses $\mathrm{\texttt{ClusterCentroid}}
      _i$; number of vertices for each elliptical fracture; number 
      of Latin Hypercube Samples (\texttt{NumLHS}); Latin Hypercube 
      Samples for fracture parameters and fracture normals (obtained 
      from Algorithm \ref{Algo:DFN_Orientation}); observation points; 
      observation pressure datasets; and flow parameters for 
      \textsf{PFLOTRAN} simulator.
    \FOR{$i = 1, 2, \cdots, \mathrm{\texttt{NumLHS}}$}
      \IF {\textbf{Case \#1:}~Constant aperture and constant permeability 
        for all the fractures}
        \STATE LHS of each fracture includes:~Fracture normal based on strike 
          and dip angle, minor axis radius, and aspect ratio (length of major 
          axis to minor axis). All these parameters are sampled from uniform 
          distributions.
        \STATE LHS, which are the same for all fractures:~Aperture and 
          permeability, whose log-values are sampled from uniform distributions.
      \ENDIF
      \IF {\textbf{Case \#2:}~Fracture aperture sampled from log-normal distribution}
        \STATE LHS of each fracture includes:~Fracture normal based on strike 
          and dip angle, minor axis radius, aspect ratio of each fracture, and 
          standard deviation for log-normal aperture distribution. All these 
          parameters are sampled from uniform distributions.
      \ENDIF
      \IF {\textbf{Case \#3:}~Fracture aperture from fracture transmissivity}
        \STATE LHS of each fracture includes:~Fracture normal based on strike 
          and dip angle, minor axis radius, aspect ratio of each fracture, and 
          $\alpha$. All these parameters are sampled from uniform distributions.
      \ENDIF
      \IF {\textbf{Case \#4:}~Length correlated aperture}
        \STATE LHS of each fracture similar to \textbf{Case \#3}.
      \ENDIF
      \STATE For each observation point, find the closest mesh cell 
        and extract the model pressure from this cell.
      \STATE Calculate the misfit functional values given by equation 
        \eqref{Eqn:Misfit_Functional}. 
      \STATE Choose the fracture parameters and normals, which has 
        the minimum misfit functional value.
    \ENDFOR
    \STATE OUTPUT (for each case):~Fracture lengths $l^*_i$ 
      (major axis length, minor axis length, and mean length), 
      unit normals $\mathbf{\hat{n}}_i$, apertures $b_i$, and 
      permeabilities $k_{s,i}$, where $i = 1, 2, \cdots, n$.
  \end{algorithmic}
\end{algorithm}

Once, we obtain the constrains on fracture orientation, and  if 
we know the bounds on fracture length, we can estimate 
the fracture lengths by minimizing the misfit between flow measurement 
data and model data. Bounds on the fracture length can be 
obtained from either microseismic data through fracture compliance and 
fracture length relationship \cite{2013_Verdon_Wustefeld_GP_v61_p461_p475} 
or from bounding relationships for hydraulic fracture height growth to 
hydraulic fracture fluid volume \cite{2013_Flewelling_etal_v40_p3602_p3606}.
Correspondingly, the misfit objective functional to be minimized is given 
by the following equation: 
\begin{align}
  \label{Eqn:Misfit_Functional}
  \mathcal{J}&[(\theta^{*}_1, \phi^{*}_1, l^{*}_1), 
  (\theta^{*}_2, \phi^{*}_2, l^{*}_2), \cdots, 
  (\theta^{*}_n, \phi^{*}_n, l^{*}_n)] = \nonumber \\
  &\displaystyle \sum \limits_{i = 1}^{\mathrm{\texttt{NumObs}}} 
  \frac{1}{p_{\text{\tiny {maxobs}}}} \left(p_i((\theta_1, 
  \phi_1, l_1), (\theta_2, \phi_2, l_2), \cdots, (\theta_n, 
  \phi_n, l_n), \mathbf{x},t) - p_{\text{\tiny {obs}},i} \right)^2
\end{align}
where $n$ is the number of fractures in the DFN. $\theta_{\mathfrak{e}}$ 
is the strike angle, $\phi_{\mathfrak{e}}$ is the dip angle, and $l_
{\mathfrak{e}}$ is the fracture length, where $\mathfrak{e} = 1, 2, 
\cdots, n$. \texttt{NumObs} is the total number of observation points.
$p_{\text{\tiny {obs}},i}$ is the pressure data $[\mathrm{Pa}]$ at 
$i$-th observation point. $p_{\text{\tiny {maxobs}}}$ correspond to 
the maximum value of observation pressure. $p_i$ is the model pressure 
at $i$-th observation point. Algorithm \ref{Algo:DFN_Length} provides 
a detailed methodology to constrain fracture length using HPC toolkits
\textsf{dfnWorks} and \textsf{MADS}.

Analysis is performed for four different cases of fracture 
aperture distributions (note that all fractures 
considered here are ellipses), which are described as 
follows:
\begin{description}
  \item[Case \#1] Constant fracture aperture `$b$' [m] and constant 
    fracture permeability `$k_s$' for all fractures regardless of 
    their size and location. That is, it is assumed that fracture 
    length, fracture transmissivity, and permeability are not dependent 
    on fracture aperture. For this case, there is no correlation between 
    fracture geometric properties (aperture and length) and fracture 
    material properties (transmissivity and permeability). The other 
    three cases given below assume a relationship between geometric 
    and material properties.
  \item[Case \#2] It is assumed that fracture permeability is a 
    function of fracture aperture. However, no assumption is been 
    made relating fracture permeability or fracture aperture to 
    that of the fracture length. Fracture aperture value `$b$' 
    is sampled from log-normal distribution (with a specified 
    mean and standard deviation). Permeability of each fracture 
    is a function of fracture aperture, which is given by $k_s = 
    \frac{b^2}{12}$ \cite[Chapter-4]{Adler_etal}.
  \item[Case \#3] For this case, it assumed that there is a strong
    correlation between fracture aperture, fracture transmissivity, 
    and fracture permeability. Fracture aperture `$b$' is constructed 
    from fracture transmissivity $\varsigma = \mathfrak{F} (0.5l_{\text{\tiny 
    {mean}}})^{\alpha}$ by the following relation: $b = \left(\frac{12 
    \varsigma \mu}{\rho g} \right)^{1/3}$. The constants $\mathfrak{F}$ 
    and $\alpha$ depend on the underlying rock properties \cite[Chapter-4]
    {Adler_etal}. $l_{\text{\tiny {mean}}}$ corresponds to the mean 
    fracture length, which is the average of major and minor lengths 
    of the underlying elliptical fracture. For this case, note that 
    the fracture permeability $k_s = \frac{b^2}{12}$ depends on the 
    length of the fracture. 
  \item[Case \#4] Length correlated aperture, where fracture aperture 
    is defined as a function of mean fracture length by the following 
    expression: $b = \mathfrak{F}_l (0.5l_{\text{\tiny {mean}}})^{\alpha}$. 
    The constant $\mathfrak{F}_l$ depends on the underlying rock 
    properties (note that $\mathfrak{F}_l$ need not be equal to 
    $\mathfrak{F}$). Similar to the previous case, the fracture 
    permeability depends on the fracture length, which has profound 
    influence on resulting model pressure and flow rates.
\end{description}
The logic behind considering the four different cases is as follows:~In 
order to constrain fracture length from fluid flow measurements, 
it is important to take into account various correlations that exist 
between fracture geometric properties and material properties. This 
is because fluid flow measurements can only provide information related 
to pressure and velocity of the fluid flow in fractures. While laboratory 
experiments on rock cores samples drilled from 
a reservoir site provide local information of the reservoir such as local 
fracture aperture, local fracture transmissivity, and local fracture 
permeability (based on which an estimate on fracture length of the core 
sample can be obtained). However, there are uncertainties associated in 
extrapolating the findings of the laboratory experiments on rock core 
samples to reservoir (which is at a global scale). Hence, we plan to 
investigate the above four different cases, which are of practical 
interest in fracture flow applications.

\subsection{Logic behind the proposed approach}
\label{SubSec:S2_JI_Framework}
\begin{figure}
  \includegraphics[scale = 0.74]{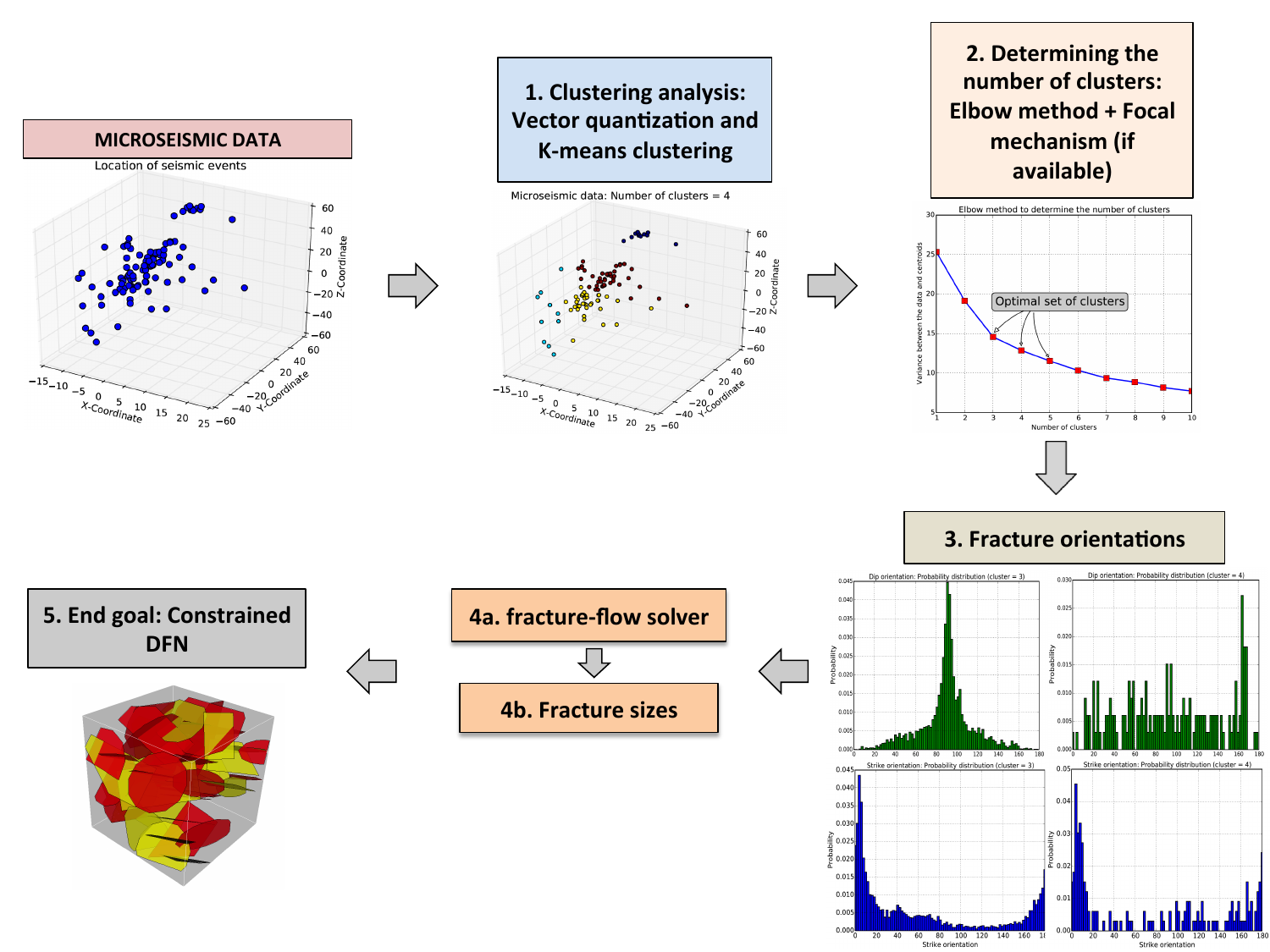}
  \caption{\textsf{\textbf{Joint geophysical and flow inversion framework 
    workflow diagram:}}~First, the microseismic data (which 
    is the coordinates of the seismic events) are clustered. The 
    number of clusters is determined based on a combination of 
    elbow method and focal mechanisms. Clustering analysis is 
    performed using $k$-means clustering algorithm. Following 
    this, a probability distribution for strike and dip angles 
    is obtained for each cluster. In combination with focal 
    mechanisms, these discrete probability distributions for
    strike and dip angles provide information on major fault/fracture 
    orientations. In addition, clustering analysis can provide 
    (possible) bounds on other fracture statistics \cite{1993_Sahimi_etal_PRL_p2186_p2189,
    2005_Dzwinel_etal_NPG_v12_p117_p128,2008_Zaliapin_etal_PRL_v101_p018501}, 
    which will be considered in our future works. Second, based on 
    these fracture orientations and \textit{a priori} information on the ranges 
    for fracture lengths, a sequence of DFNs are 
    constructed. The parameters for the DFNs are sampled from a given 
    set of probability distributions based on Latin hypercube sampling. 
    The governing equations for the flow are solved on these DFNs. 
    Finally, a set of constrained DFNs are obtained by minimizing 
    the misfit functional based on a given set of pressure observations.
  \label{Fig:JI_Workflow_Diagram}}
\end{figure}

The overall workflow for constructing the sequential
inversion framework based on geophysical and flow data is summarized 
in Figure \ref{Fig:JI_Workflow_Diagram}.
\begin{description}
  \item[Why sequential inversion?]~Sequential inversion is 
    a multi-step process. This procedure gives the flexibility 
    to invert for system parameters one at a time by considering 
    single data stream among multiple data streams. In joint 
    inversion, one inverts for the all system parameters in 
    one go by considering multiple data streams all at a time. 
    An advantage of sequential inversion procedure is that one 
    can couple existing simulators or codes, tailored to invert 
    a specific data stream. In this paper, we first invert 
    geophysical data to get fracture orientation distribution. 
    Then, we invert flow data to get fracture lengths.
  \item[Why need cluster analysis?]~Clustering analysis 
    of inverted microseismic events can provide: (i) probability 
    distribution of possible fracture orientations and (ii) 
    fracture plane centroids, which are the individual cluster 
    centers. However, in data clustering analysis, determining 
    the number of clusters/fractures is a frequent problem 
    \cite{1985_Milligan_Cooper_Psy_v50_p159_p179,
    2003_Sugar_James_JASA_v98_p750_p763,
    2007_Yan_Ye_Biometrics_v63_p1031_p1037,
    2010_He_etal_IEEETPAMI_v32_p2006_p2021,
    2010_Wang_Biometrika_v32_p2006_p2021}, 
    and is a distinct issue from the process of actually 
    solving the clustering problem. The correct choice of 
    number of clusters is often ambiguous, and is dependent 
    on the desired clustering resolution of the user 
    \cite{1953_Thorndike_Psy_v18_p267_p276,
    2010_Chiang_Mirkin_JC_v27_p2_p40}. 
    For example, in $k$-means clustering, increasing $k$ 
    without penalty will always reduce the amount of error 
    in the resulting clustering, to the extreme case of zero 
    error if each data point is considered its own cluster 
    (i.e., when $k$ equals the number of data points, 
    $n$). But this may not be the correct or optimal 
    choice of number of clusters. Intuitively, the optimal 
    choice of number of clusters $k$ will strike a balance 
    between maximum compression of the data using a single 
    cluster, and maximum accuracy by assigning each data 
    point to its own cluster.
  \item[How can physics enhance clustering analysis?]~In order 
    to find number of clusters $k$ in $k$-means clustering, we 
    use traditional elbow method \cite{2013_Kodinariya_Makwana_IJ_v1_p90_p95,
    2005_Tibshirani_Walther_v14_p511_p528,2001_Tibshirani_etal_v63_p411_p423}. 
    The elbow method looks at the percentage of variance explained 
    as a function of the number of clusters. Intuitively, if one 
    plots the percentage of variance explained by the clusters 
    against the number of clusters, the first cluster will add 
    much information (explains a lot of variance), but at some 
    point the marginal gain will drop, giving an angle in the 
    graph (for example see Figure \ref{Fig:JI_Entire_Cluster_1}(a) 
    and (b)). That is, at this point adding another cluster 
    doesn't give much better modeling of the data. The number 
    of clusters $k$ is chosen at this point, hence the `elbow 
    criterion'. This `elbow' cannot always be unambiguously 
    identified. In order to reduce the uncertainty associated 
    with identifying this elbow we rely on the dip and strike 
    angle distribution obtained from focal mechanism physics.
    The number of peaks (which correspond to dominant fracture 
    planes) in these fracture angle distributions narrows down 
    the uncertainty associated in determining the number of 
    clusters from elbow method. 
\end{description}
In the next section, we demonstrate the efficacy of 
this multi-physics based sequential inversion 
framework through a representative synthetic example.

%

\section{RESULTS:~A SYNTHETIC NUMERICAL EXAMPLE}
\label{Sec:S3_JI_Results}
\begin{figure}
  \centering
  \subfigure[Map view:~Distribution of microseismic events and geophones]
    {\includegraphics[scale=0.995]
    {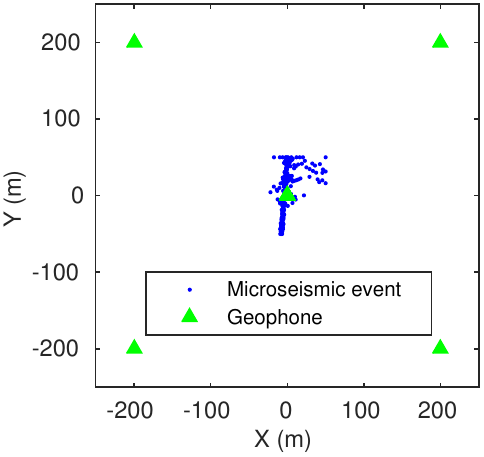}}
  \subfigure[Depth view:~Distribution of microseismic events and geophones]
    {\includegraphics[scale=0.995]
    {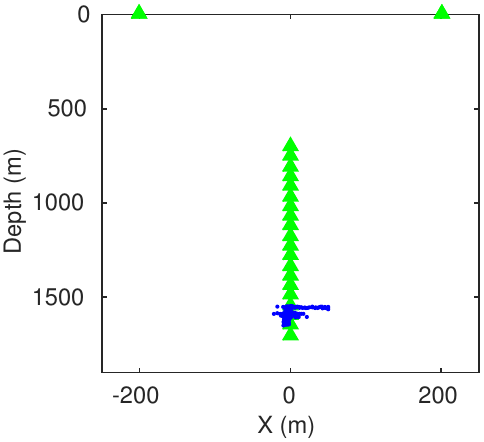}}
  \subfigure[Velocity model]
    {\includegraphics[scale=0.795]
    {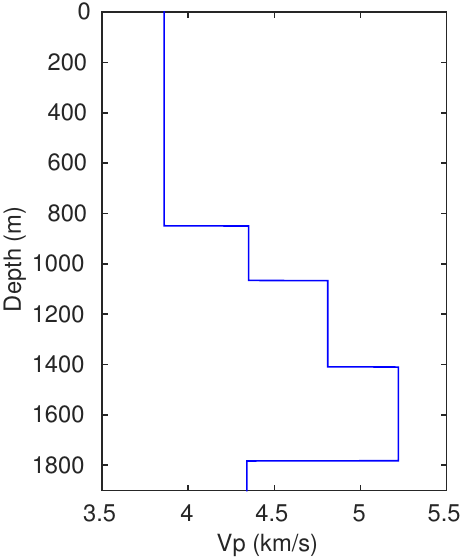}}
  \caption{\textsf{\textbf{Geophones arrangement and synthetic 
    velocity model:}}~Figures (a) and (b) provide 
    the map and depth view of the distribution of microseismic 
    events and geophones for the synthetic example. Figure (c)
    shows the 1D compressional-velocity ($\mathrm{V}_p$) 
    model used for this synthetic example. The shear-velocity 
    ($\mathrm{V}_s$) model is similar to $\mathrm{V}_p$ model, 
    with a constant, which is assumed to be: $\frac{\mathrm{V}_
    p}{\mathrm{V}_s} = 1.73$.
  \label{Fig:JI_Geophones_Microseismics}}
\end{figure}

As a synthetic example, we construct a distribution of geophones
and a one-dimensional velocity model (see Figure \ref{Fig:JI_Geophones_Microseismics}). 
A total of 4 surface geophones and 20 borehole geophones are placed 
near the fractures of interest. It is assumed that these three-component 
geophones can record both compressional and shear waves. Approximately, 
330 microseismic events are assumed to occur on the fractures of interest 
and be recorded by all geophones. Each event has its own location and 
focal mechanism. Based on this synthetic velocity model, we first invert 
for microseismic event locations using seismic wave arrival times. The 
used arrival times are true arrival times modified by adding a Gaussian 
distribution of 2 ms noise to simulate observation errors. With observation 
errors, the inverted microseismic events are not located exactly on the 
fractures, but scattered around. To locate fracture centers, 
$k$-means clustering analysis is used. The cluster centers coordinates 
correspond to fracture plane center coordinates.

The size of the domain of interest is a cube of $200 \times 200 \times 
200 \; [\mathrm{m}^3]$. The reference datum for vertical depth is at 
1500 m, which is the top surface of the cube. We assume that there 
are three major fractures in the subsurface system, whose unit normals 
are given by:~$\mathbf{n}_1 = -0.355 \mathbf{\hat{e}}_x - 0.646 
\mathbf{\hat{e}}_y + 0.676 \mathbf{\hat{e}}_z$, $\mathbf{n}_2 = 
-0.996 \mathbf{\hat{e}}_x + 0.077 \mathbf{\hat{e}}_y - 0.038 
\mathbf{\hat{e}}_z$, and $\mathbf{n}_3 = 0.316 \mathbf{\hat{e}}_x 
+ 0.715 \mathbf{\hat{e}}_y + 0.623 \mathbf{\hat{e}}_z$. In terms 
of strike and dip angles, these are close to $(5^{\mathrm{o}}, 
90^{\mathrm{o}})$, $(117^{\mathrm{o}}, 60^{\mathrm{o}})$, and 
$(120^{\mathrm{o}}, 120^{\mathrm{o}})$. The major fractures 
are assumed to be ellipses with centers located at $(-5.543, 
-19.861, 98.218)$, $(0.577, 19.39, 91.1)$, and $(9.42, 39.088, 
53.548)$, where the reference datum coordinates (which is equal
to $(0, 0, 1500)$) are subtracted to put these coordinates in 
the domain of interest (for instance, see Figure \ref{Fig:JI_DFN_Pressure_TrueSoln}).
It is assumed that the length of each elliptical fracture in 
the direction of minor axis is equal to 250 m. The major axis 
length of ellipse is assumed to be equal to 275 m, 300 m, and 
312.5 m. The mean length for these three fractures are 262.5 m, 
275 m, and 281.25 m. As the fracture sizes are greater than the 
domain size of interest, most of them are truncated to fit in 
the specified dimensions of the cube. 

For simplicity and to demonstrate various ideas of the proposed 
sequential inversion framework, steady-state analysis is performed 
for the flow problem on DFN. The following 
boundary conditions, fracture parameters, and flow parameters 
are assumed: 
\begin{itemize}
  \item \textbf{Flow boundary conditions:}~Dirichlet boundary 
    conditions are prescribed on the left and right sides of the 
    cube. Pressure on the left side of the cube is equal 
    to 30 MPa while that on the right side is equal to 10 MPa. 
    On all other sides of the domain, the flux is set to zero.
  \item \textbf{Flow parameters:}~Constant fracture aperture is 
    assumed, which is taken as $10^{-5} \; [\mathrm{m}]$. Correspondingly, 
    constant fracture permeability is taken to be equal to $10^{-12} 
    \; [\mathrm{m}^2]$. Other fluid and porous media parameters, 
    based on the properties of 
    water and fractured granite, are set to:
     $\varphi = 0.25$, $\rho = 997 \; [\mathrm{Kg} \, 
    \mathrm{m}^{-3}]$, $Q_w = 0$, $\mu = 8.94 \times 10^{-4} \; 
    [\mathrm{Pa} \, \mathrm{s}]$, and $g = 9.8 \; [\mathrm{m} \, 
    \mathrm{s}^{-2}]$. 
\end{itemize}

Using these parameters, the fluid pressure profile (which 
is the true solution) is obtained on the original DFN, 
shown in Figure \ref{Fig:JI_DFN_Pressure_TrueSoln}. Pressure 
observation data, used in our inversion framework,
is sampled at the centroids of the elliptical 
fractures, whose values are given as follows:
\begin{itemize}
  \item \textbf{Pressure observation data:}~$p_
    {\text{\tiny {obs}},1} = 21.86$ MPa, $p_
    {\text{\tiny {obs}},2} = 19.08$ MPa, and 
    $p_{\text{\tiny {obs}},3} = 18.33$ MPa.
\end{itemize}

\subsection{Discussion of numerical results and inferences}
\label{SubSec:Discussion}
Using the above observational pressure datasets, inverted 
seismic events locations and focal mechanisms, we perform 
sequential inversion analysis to determine the fracture characteristics 
for various cases discussed in subsection \ref{SubSec:S2_JI_Framework}. 
The following are the numerical values for the input parameters 
for Algorithms \ref{Algo:DFN_Orientation} and \ref{Algo:DFN_Length}:
\begin{itemize}
  \item \texttt{NumCoord} = 332, \texttt{MaxIters} = 20, \texttt{TolValue} 
  = $10^{-8}$, \texttt{MaxCombo} = $10^7$
  \item The Monte Carlo simulations
    based on LHS are constructed
    using the following ranges with \texttt{NumLHS} = 10:\\
    Elliptical fracture lengths (in the direction 
    of minor axis) are chosen to be between 200 m and 300 m.  The 
    aspect ratio (major axis length to minor axis length) is varied from 
    1.0 to 1.5. 
  Based on the case, fracture aperture and permeability ranges are: 
  \begin{description}
    \item[Case \#1] Log of fracture aperture and log of fracture permeability 
      are varied from -6.0 to -4.0 and -13.0 to -11.0.
    \item[Case \#2] Log of mean value of aperture is -5.0 while its standard 
      deviation is varied from 0.6 to 0.9.
    \item[Case \#3] $\mathfrak{F} = 1.6 \times 10^{-9}$ and $\alpha$ is varied 
      from 0.5 to 2.0
    \item[Case \#4] $\mathfrak{F}_l = 5 \times 10^{-5}$ and $\alpha$ is varied 
      from 0.5 to 2.0
  \end{description}
\end{itemize}

\begin{remark}
  Numerical simulations are performed on using 32 cores
  to construct the strike angle and the dip angle 
  discrete probability distributions for the entire seismic events 
  using pure clustering analysis. It should be noted that the total 
  number of possible fracture orientation combinations for 332 seismic 
  events is equal to 6,044,060, which is quite high. To reduce the 
  computational time to calculate the entire discrete probability 
  distribution, we use a combination of \textsf{Multiprocessing} 
  parallelization module and \textsf{Itertools} module in \textsf{Python}. 
  Correspondingly, the computational time for these 6 million combinations 
  is around 904 seconds.
\end{remark}

\begin{figure}
  \centering
  \subfigure[True events (with Gaussian white noise)]
    {\includegraphics[scale=0.8]
    {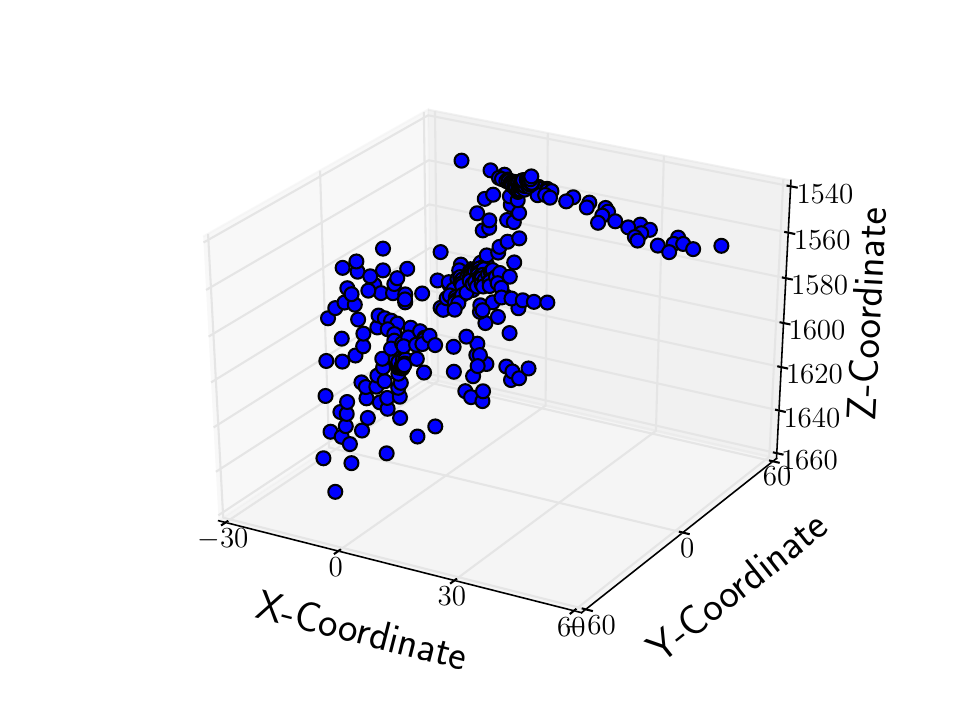}}
  \subfigure[Inverted events with errors]
    {\includegraphics[scale=0.8]
    {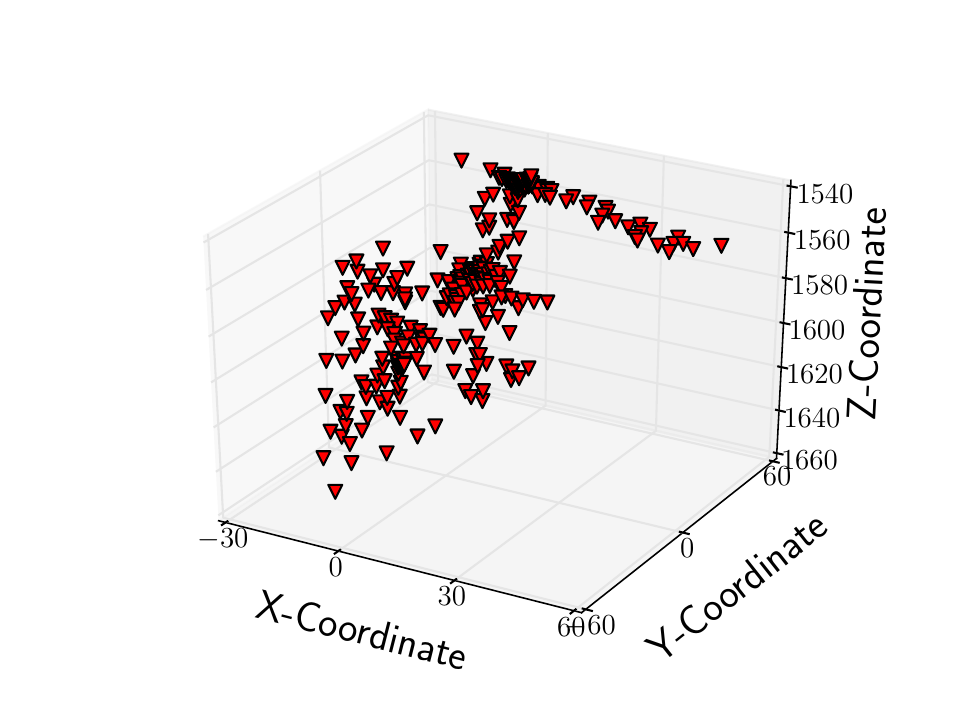}}
  \caption{\textsf{\textbf{Event locations (synthetic 
    data):}}~Figures (a) and (b) 
    show the location of true and inverted seismic 
    events, respectively.
  \label{Fig:JI_Ground_Truth_1}}
\end{figure}

\begin{figure}
  \centering
  \subfigure[Strike angle distribution (true events)]
    {\includegraphics[scale=0.4]
    {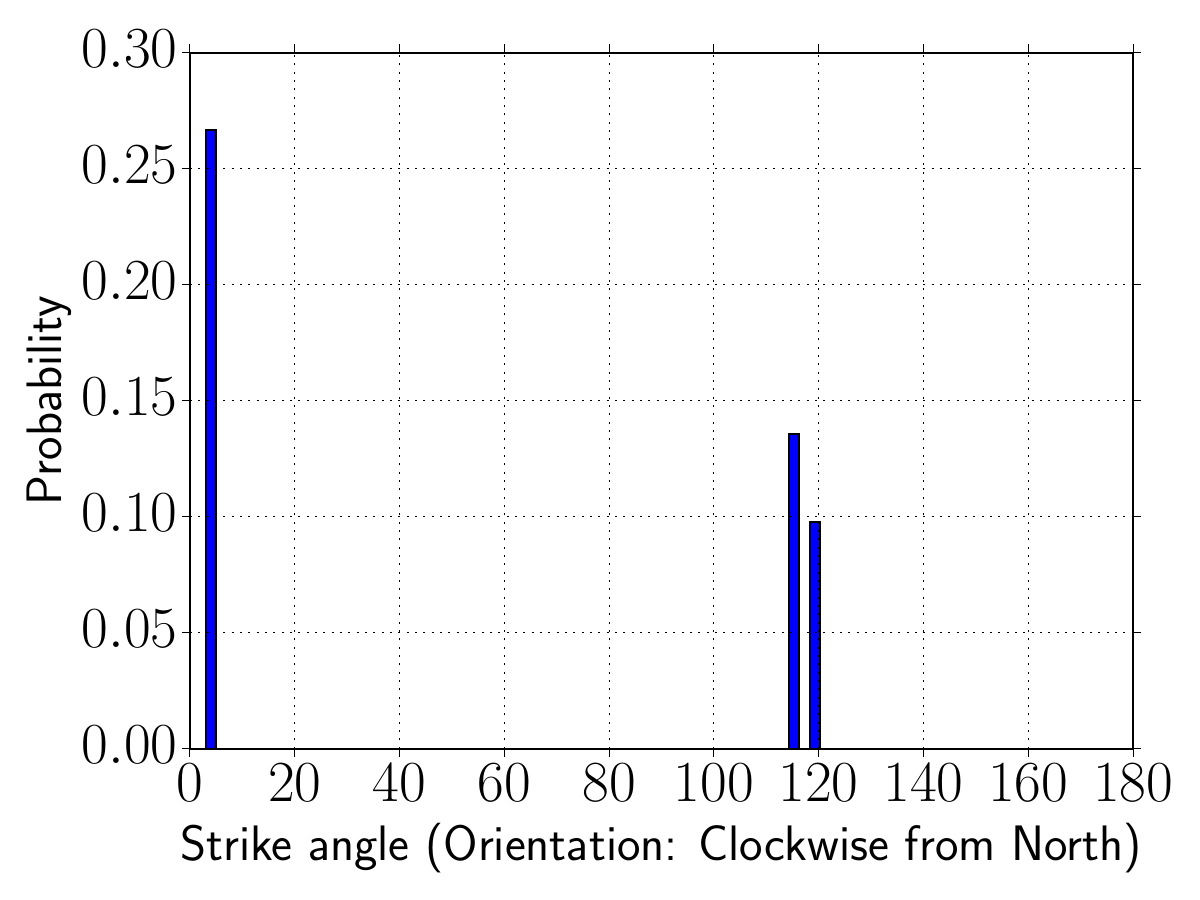}}
  \subfigure[Strike angle distribution (inverted events)]
    {\includegraphics[scale=0.4]
    {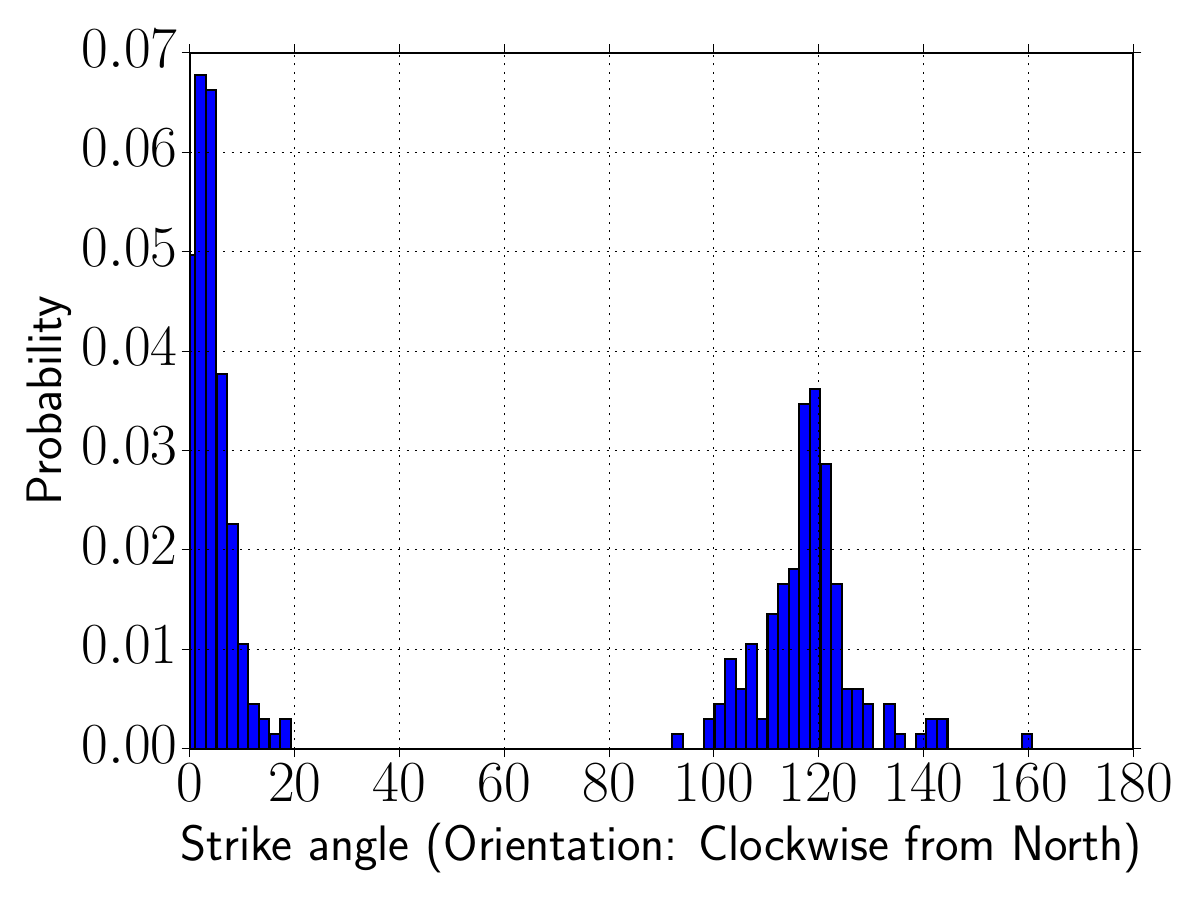}}
  \subfigure[Dip angle distribution (true events)]
    {\includegraphics[scale=0.4]
    {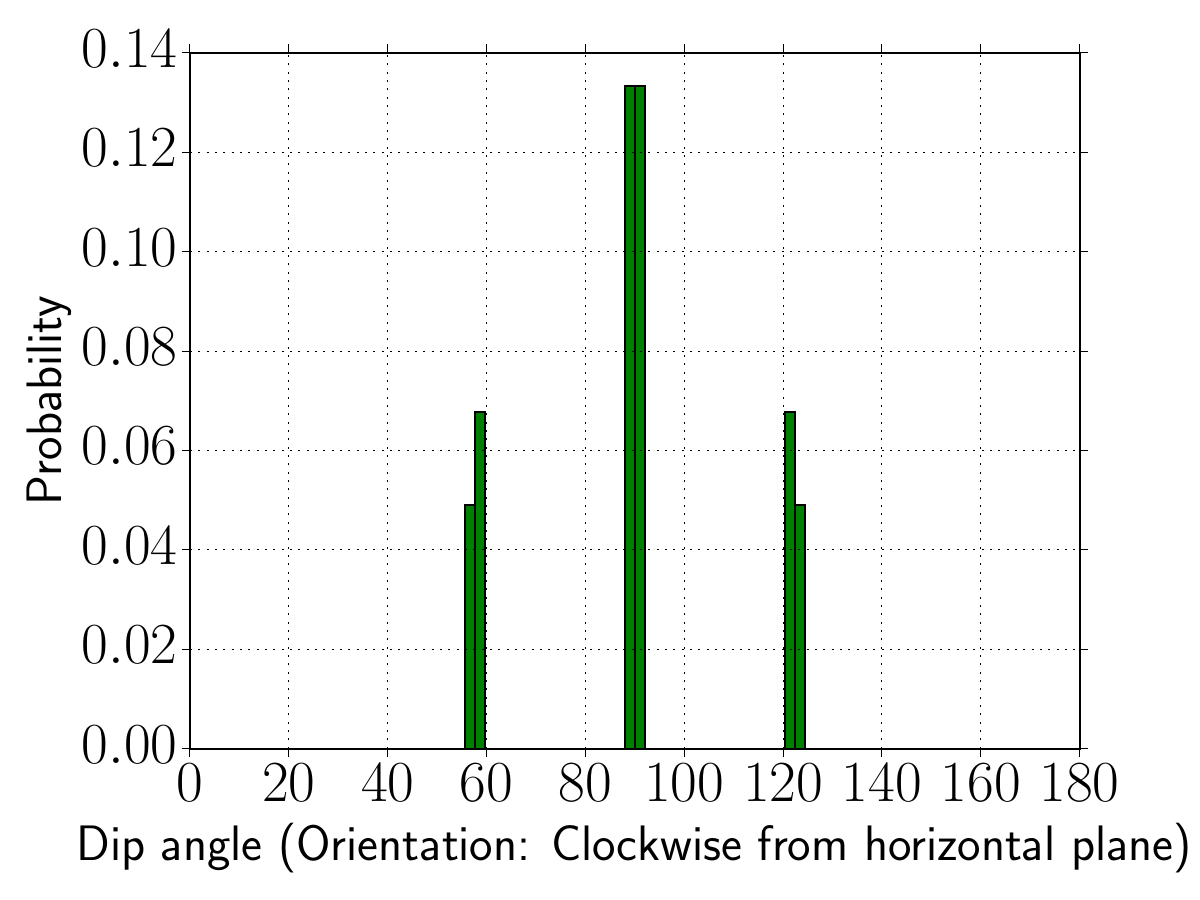}}
  \subfigure[Dip angle distribution (inverted events)]
    {\includegraphics[scale=0.4]
    {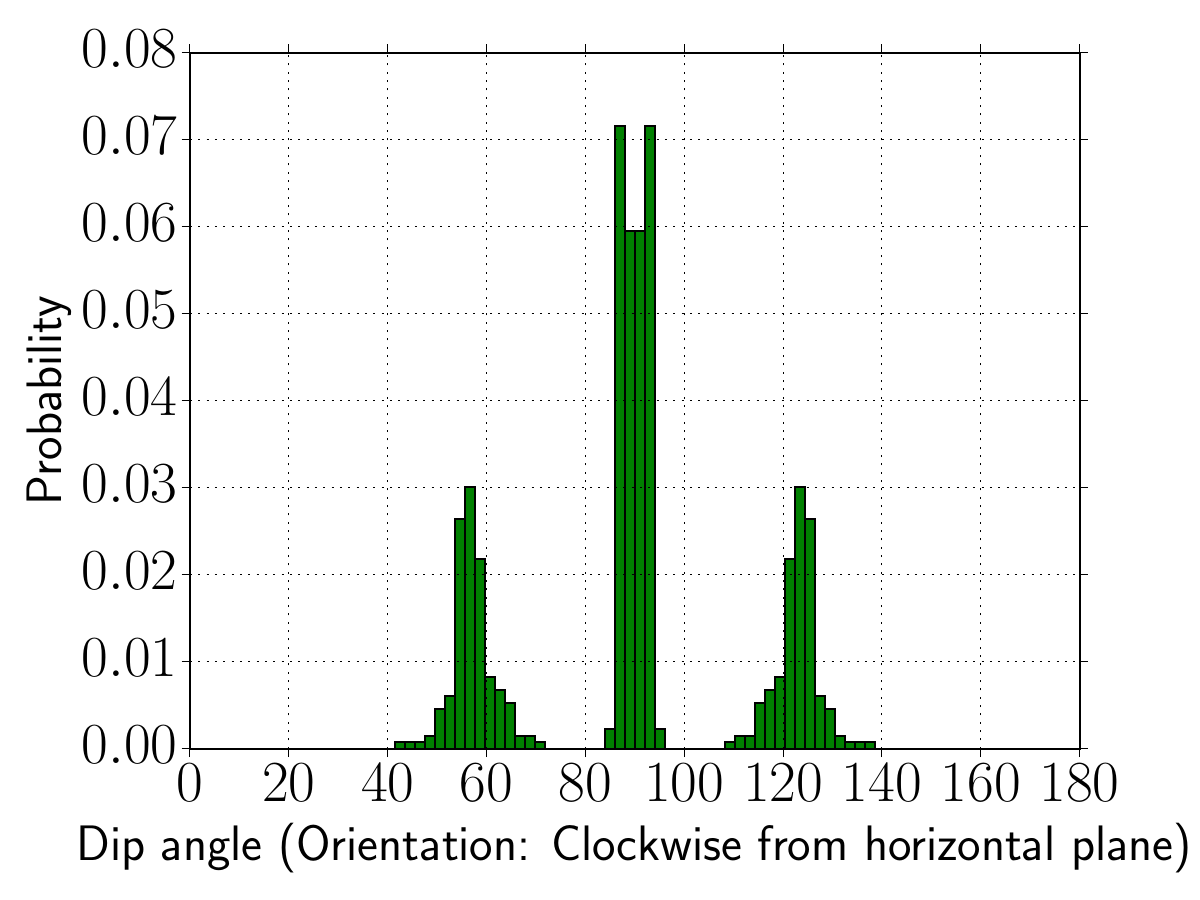}}
  \caption{\textsf{\textbf{True and inverted fault/fracture 
    orientations based on focal mechanisms:}}~Figures (a) and (b)
    show the strike angles of true and 
    inverted events while figures (c) and (d) show 
    the corresponding dip angles. From the ground 
    truth, we have three different fracture planes, 
    whose orientation in terms of the unit normals 
    are given as follows: 
    $\mathbf{n}_1 = -0.355 \mathbf{\hat{e}}_x - 0.646 
    \mathbf{\hat{e}}_y + 0.676 \mathbf{\hat{e}}_z$, 
    $\mathbf{n}_2 = -0.996 \mathbf{\hat{e}}_x + 0.077 
    \mathbf{\hat{e}}_y - 0.038 \mathbf{\hat{e}}_z$, and
    $\mathbf{n}_3 = 0.316 \mathbf{\hat{e}}_x + 0.715 
    \mathbf{\hat{e}}_y + 0.623 \mathbf{\hat{e}}_z$. 
    These correspond to strike and dip angles that 
    are close to $(5^{\mathrm{o}}, 90^{\mathrm{o}})$, 
    $(117^{\mathrm{o}}, 60^{\mathrm{o}})$, and 
    $(120^{\mathrm{o}}, 120^{\mathrm{o}})$.
  \label{Fig:JI_Ground_Truth_2}}
\end{figure}

\begin{figure}
  \centering
  \subfigure[Elbow method:~Determining the number of clusters 
    (true events)]
    {\includegraphics[scale=0.5]
    {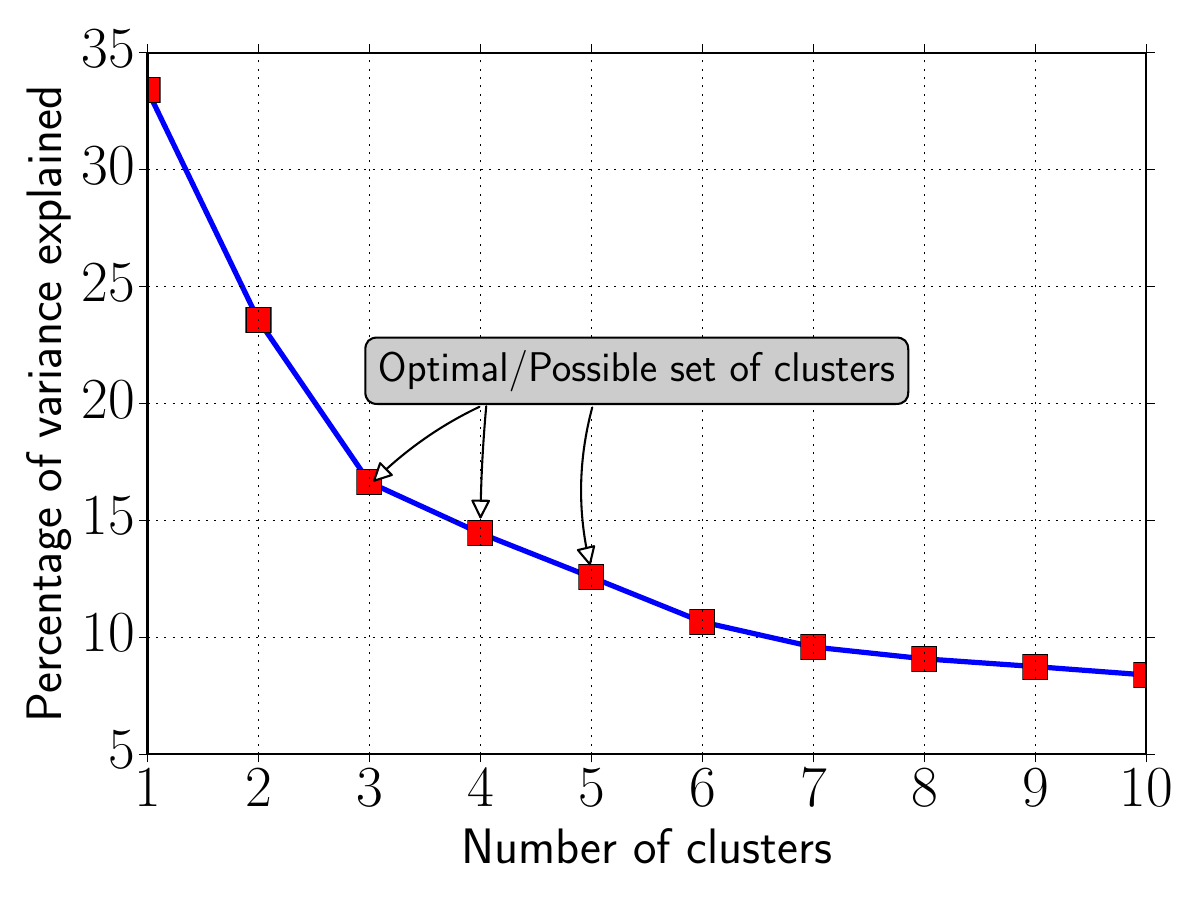}}
  \subfigure[Elbow method:~Determining the number of clusters 
    (inverted events)]
    {\includegraphics[scale=0.5]
    {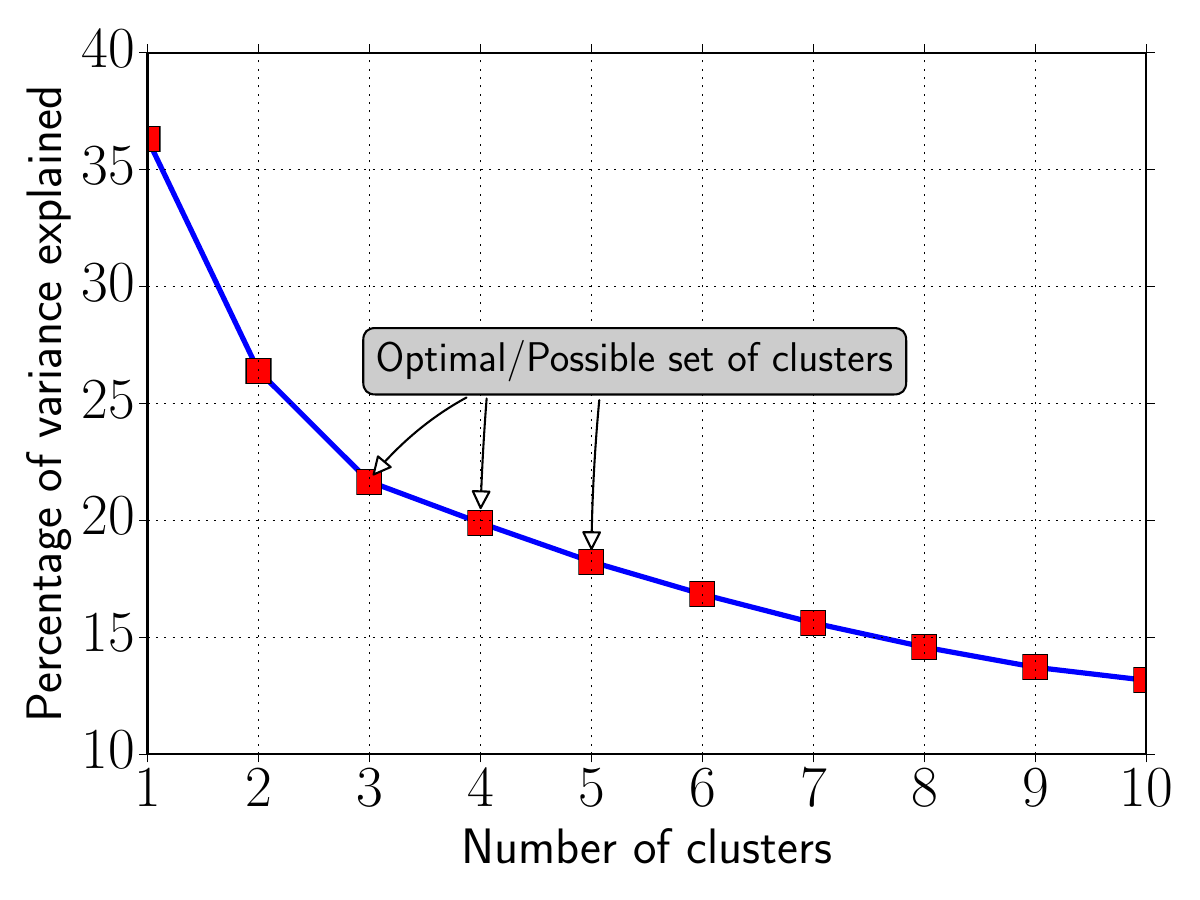}}
  \caption{\textsf{\textbf{Clustering analysis of true and 
    inverted events:}}~Figures (a) and (b)
    show the percentage of variance as 
    a function of number of clusters for true and 
    inverted events, respectively. The possible number of 
    fracture clusters based on elbow method (which is a pure 
    statistical approach) are between 3 to 5. Based on this 
    traditional clustering methodology, without the information 
    from focal mechanisms (which is a physics-based approach) it 
    should be noted that adding another cluster (beyond 5) 
    does not provide much better modeling of the seismic events.
  \label{Fig:JI_Entire_Cluster_1}}
\end{figure}

\begin{figure}
  \centering
  \subfigure[Strike angle:~Entire cluster (Ground truth)]
    {\includegraphics[scale=0.55]
    {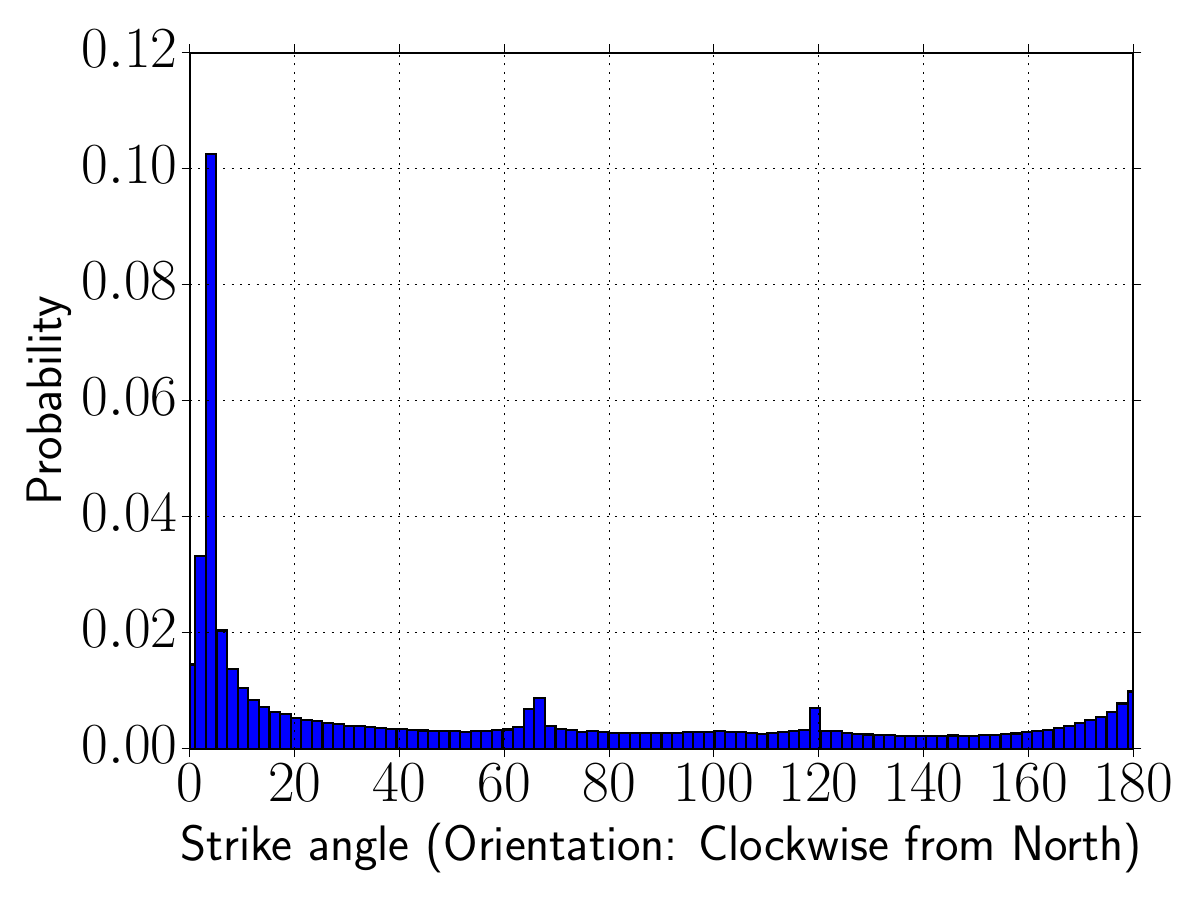}}
  \subfigure[Strike angle:~Entire cluster (Inverted events)]
    {\includegraphics[scale=0.55]
    {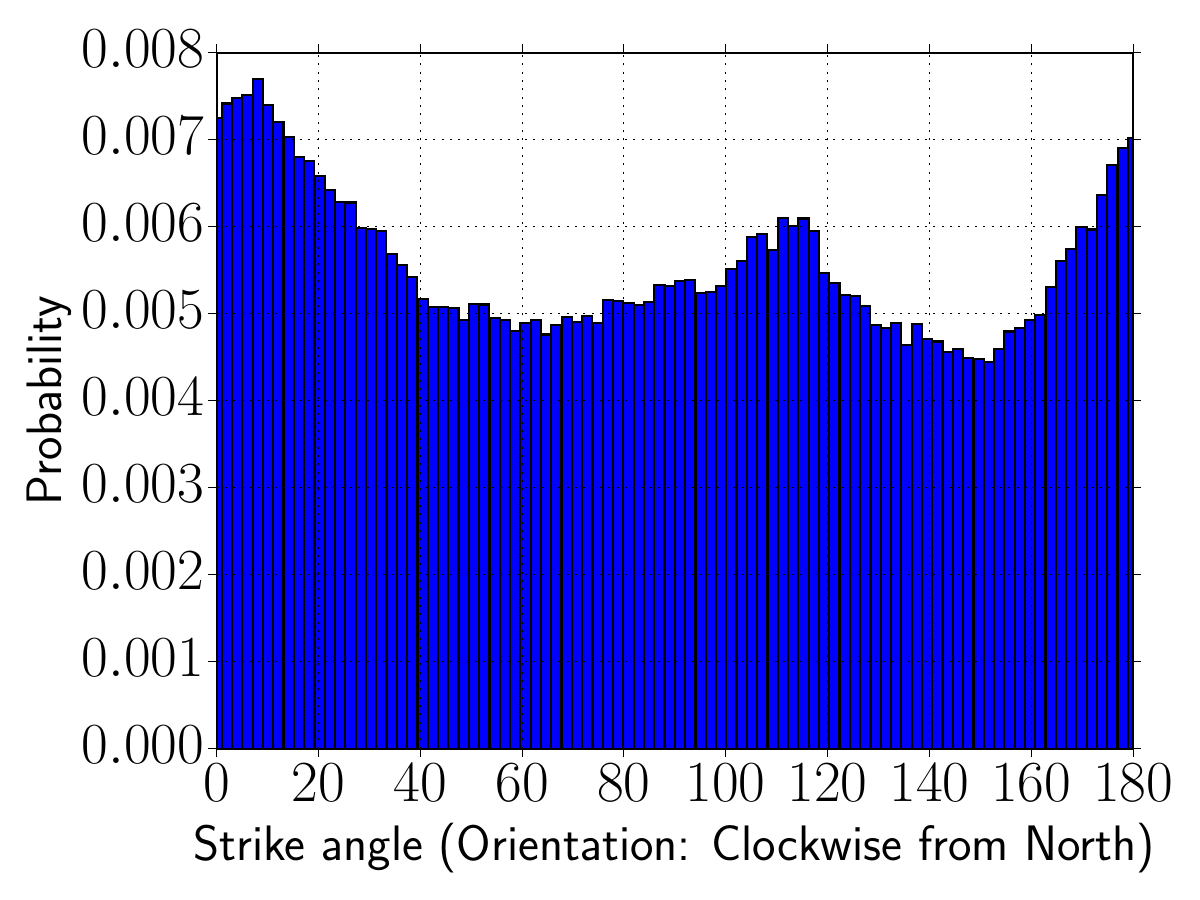}}
  \caption{\textsf{\textbf{Fault/fracture orientation of true 
    and inverted events based on clustering analysis:}}~Figures (a) 
    and (b) show the discrete probability 
    distributions for strike angle based on ground truth and 
    inverted events, respectively. Based on these figures, for inverted seismic 
    events, the dominant fracture planes have strike angles in the 
    range $0^{\mathrm{o}}-20^{\mathrm{o}}$ and $100^{\mathrm{o}}-140^{\mathrm{o}}$.
  \label{Fig:JI_Entire_Cluster_2}}
\end{figure}

\begin{figure}
  \centering
  \subfigure[Dip angle:~Entire cluster (Ground truth)]
    {\includegraphics[scale=0.55]
    {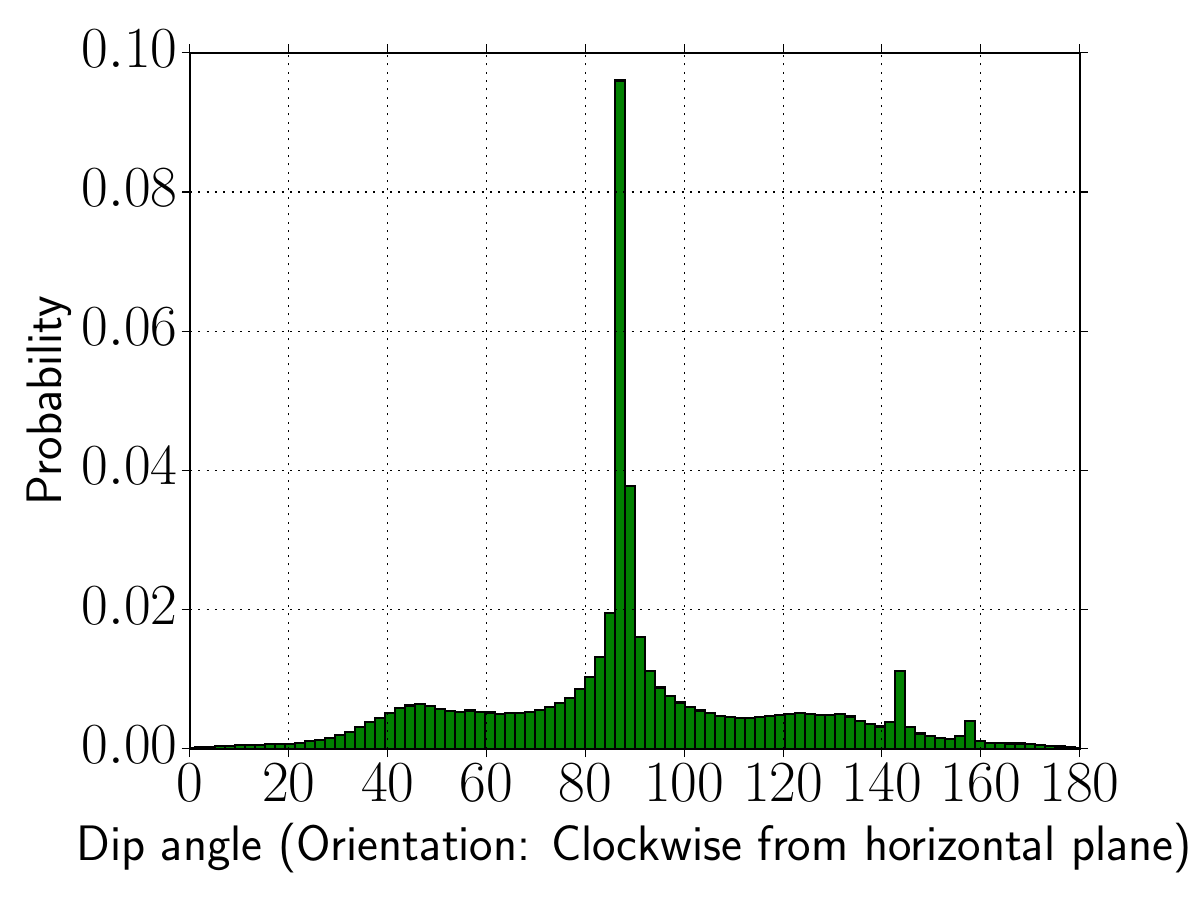}}
  \subfigure[Dip angle:~Entire cluster (Inverted events)]
    {\includegraphics[scale=0.55]
    {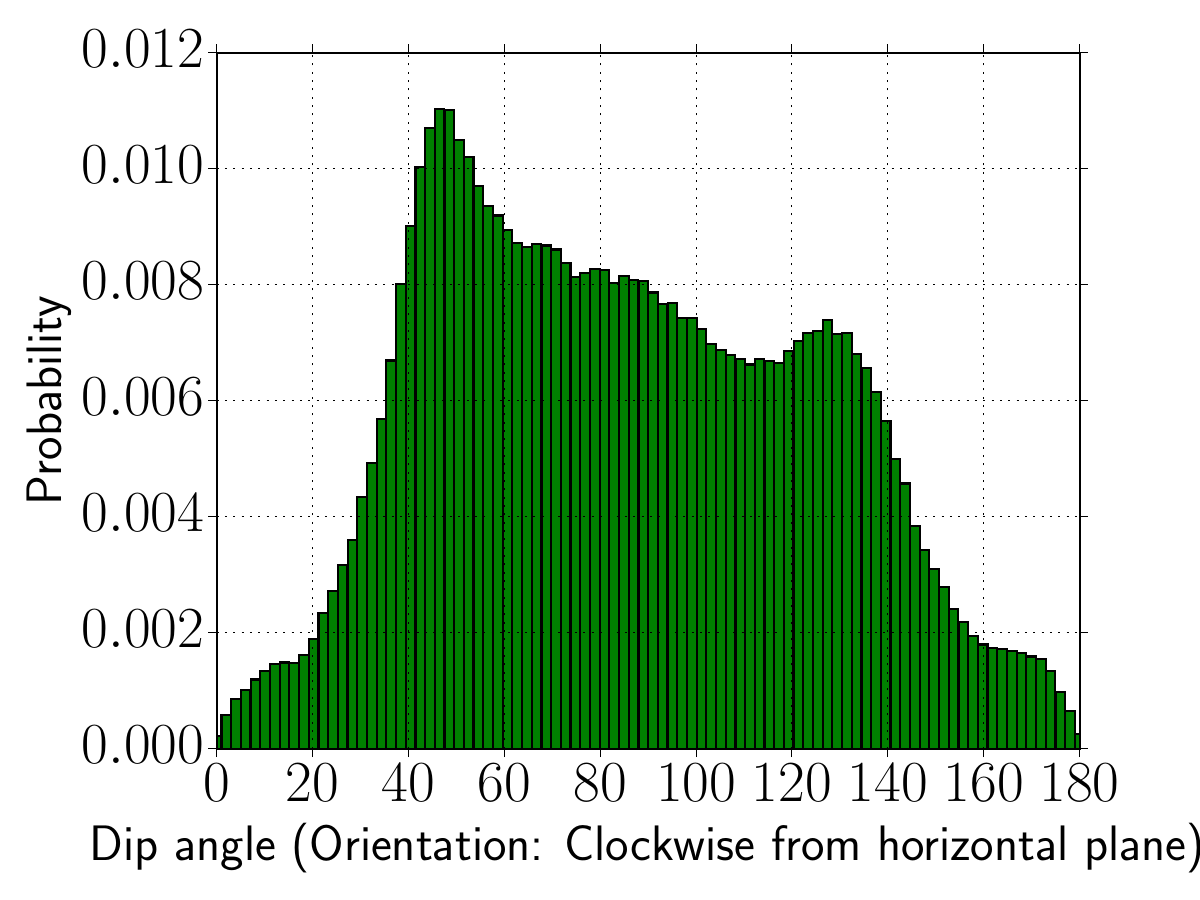}}
  \caption{\textsf{\textbf{Fault/fracture orientation of true 
    and inverted events based on clustering analysis:}}~Figures
    (a) and (b) show the discrete probability 
    distributions for dip angle angle based on ground truth 
    and inverted events, respectively. Based on these figures, for inverted 
    seismic events, the dip angles are in the range $40^{\mathrm{o}}-140^
    {\mathrm{o}}$.
  \label{Fig:JI_Entire_Cluster_3}}
\end{figure}

\begin{figure}
  \centering
  \subfigure[Strike angle:~Cluster-1 (Ground truth)]
    {\includegraphics[scale=0.45]
    {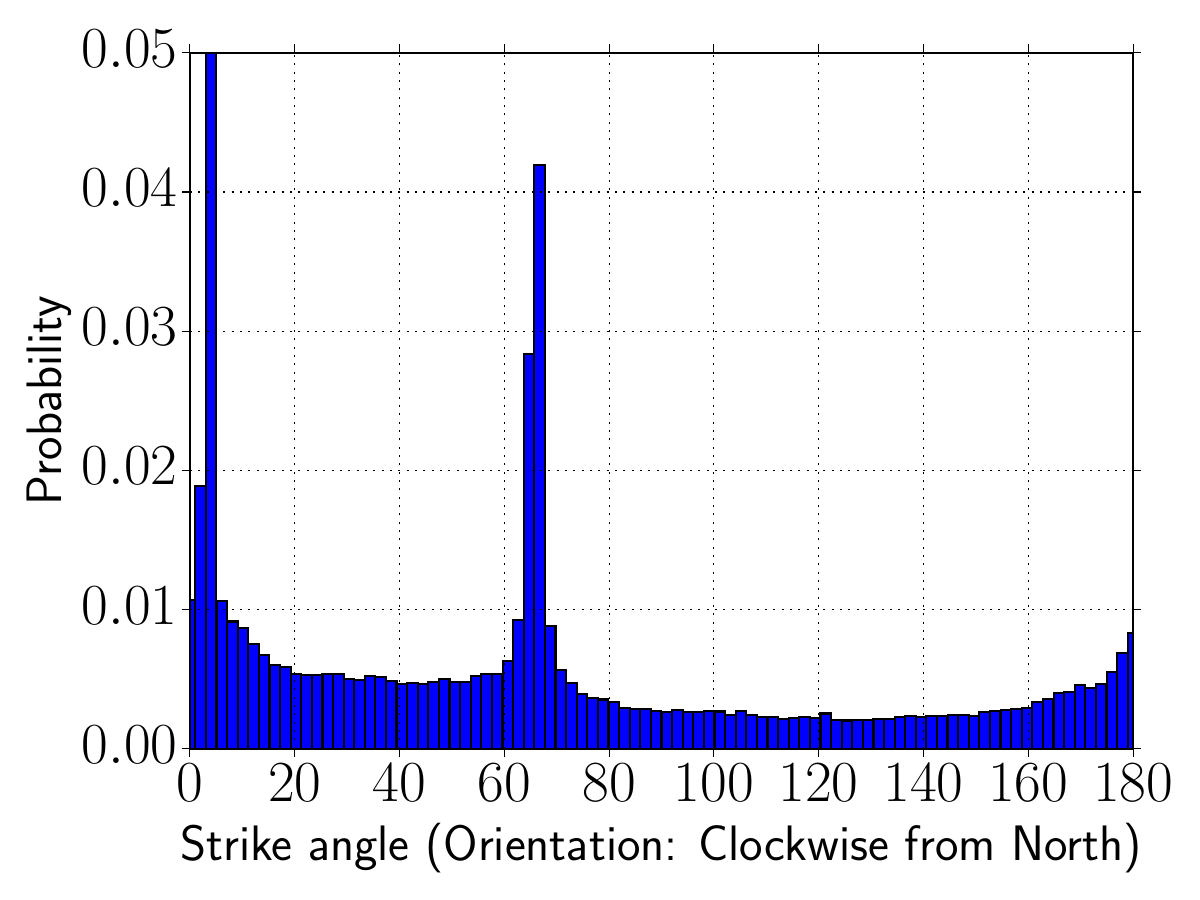}}
  \subfigure[Strike angle:~Cluster-2 (Ground truth)]
    {\includegraphics[scale=0.45]
    {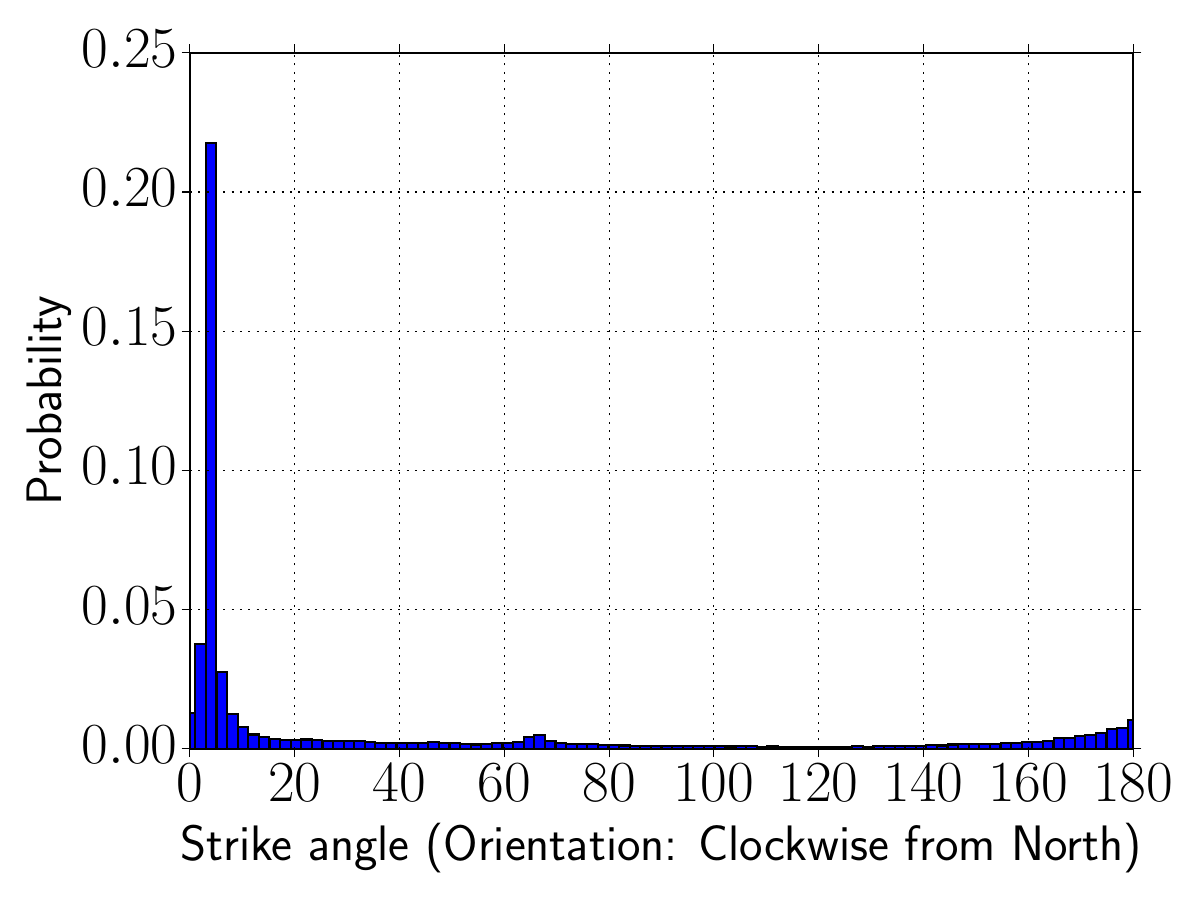}}
  \subfigure[Strike angle:~Cluster-3 (Ground truth)]
    {\includegraphics[scale=0.45]
    {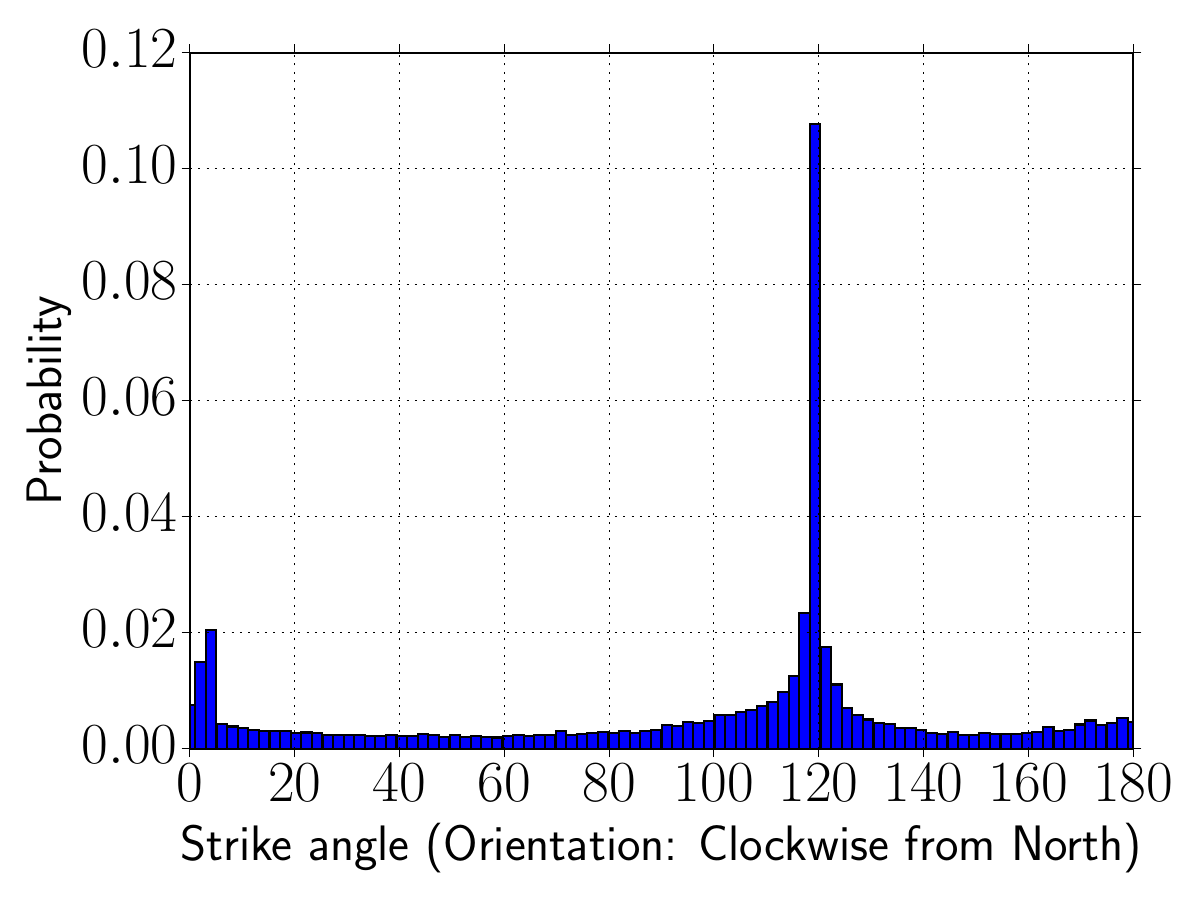}}
  \caption{\textsf{\textbf{Fault/fracture orientation (three 
    clusters):}}~Discrete probability distributions 
    for strike angle for true events based on clustering analysis.
  \label{Fig:JI_ClusteredData_True_1}}
\end{figure}

\begin{figure}
  \centering
  \subfigure[Dip angle:~Cluster-1 (Ground truth)]
    {\includegraphics[scale=0.45]
    {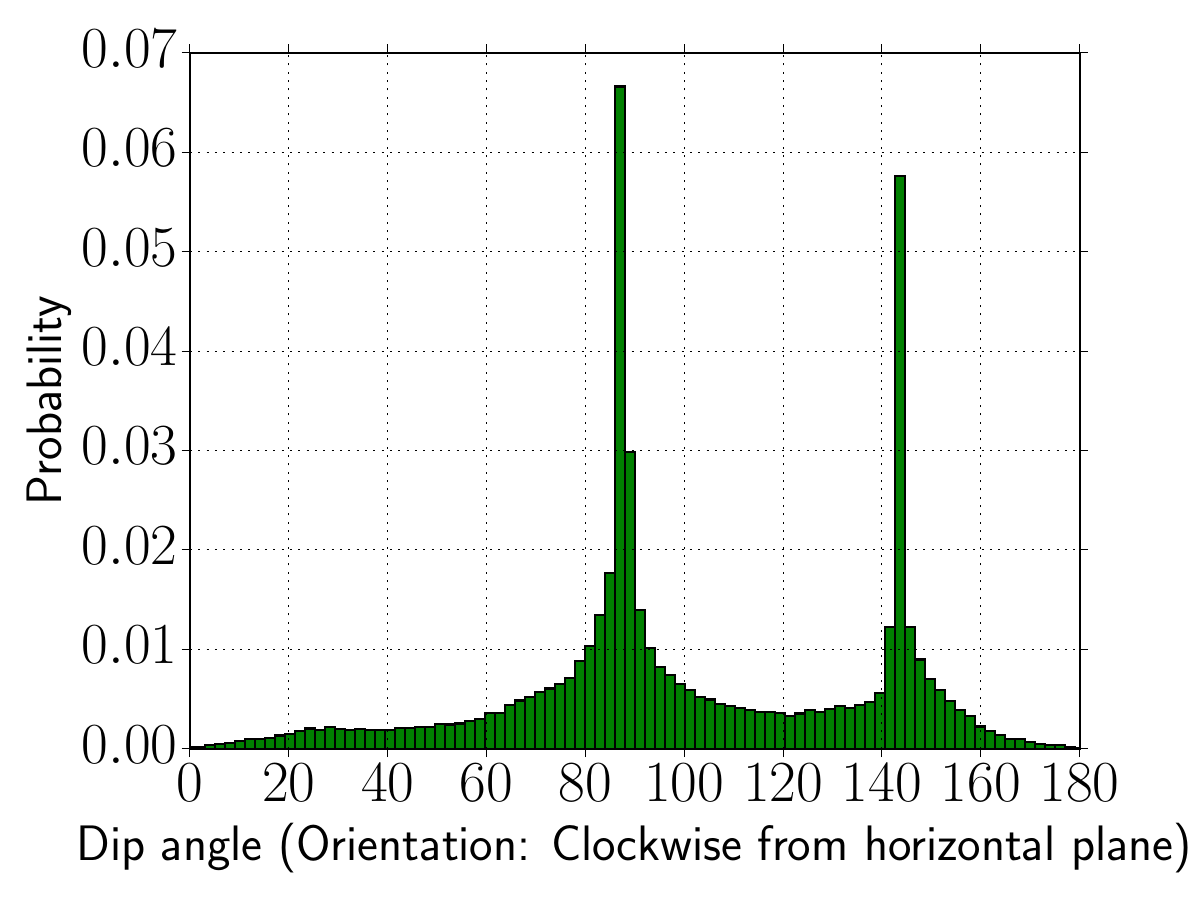}}
  \subfigure[Dip angle:~Cluster-2 (Ground truth)]
    {\includegraphics[scale=0.45]
    {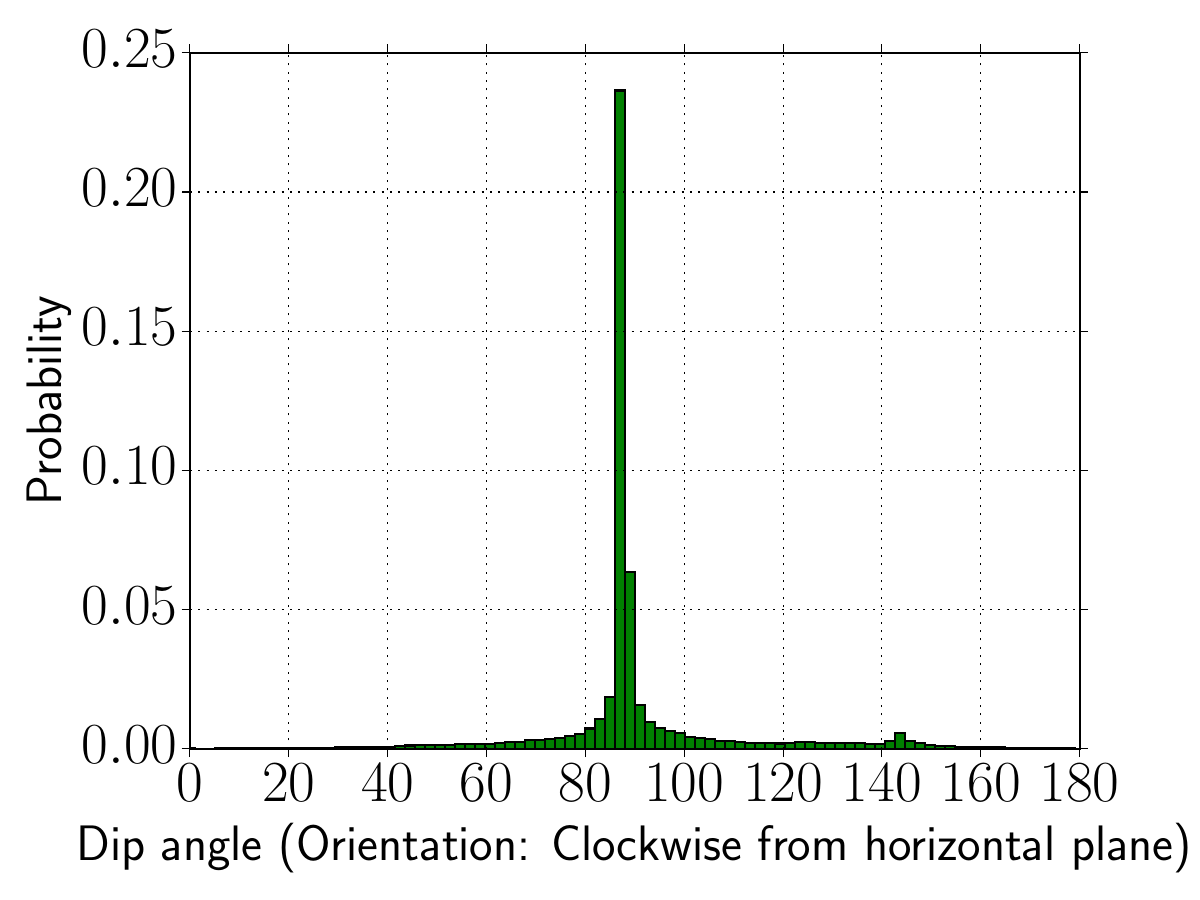}}
  \subfigure[Dip angle:~Cluster-3 (Ground truth)]
    {\includegraphics[scale=0.45]
    {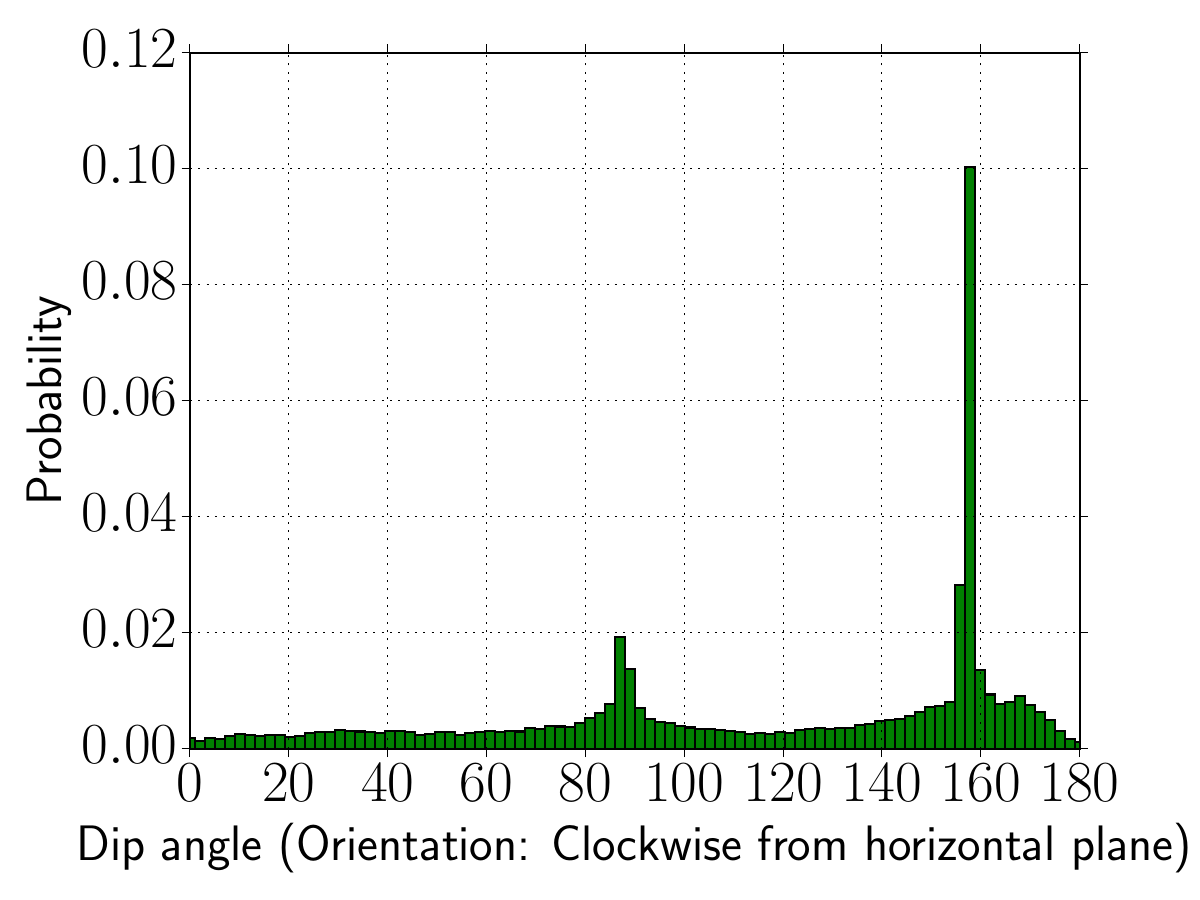}}
  \caption{\textsf{\textbf{Fault/fracture orientation (three 
    clusters):}}~Discrete probability distributions 
    for dip angle for true events based on clustering analysis.
  \label{Fig:JI_ClusteredData_True_2}}
\end{figure}

\begin{figure}
  \centering
  \subfigure[Strike angle:~Cluster-1 (Inverted events)]
    {\includegraphics[scale=0.45]
    {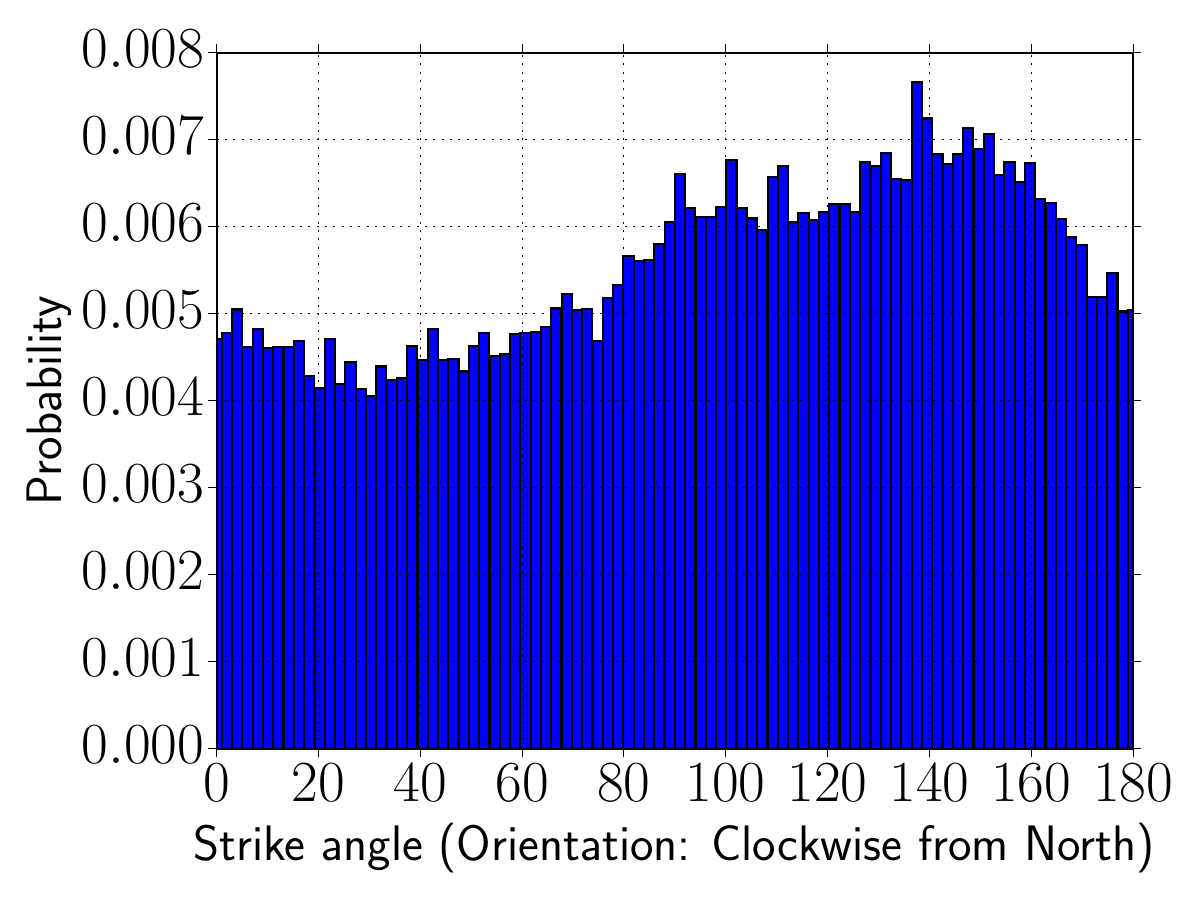}}
  \subfigure[Strike angle:~Cluster-2 (Inverted events)]
    {\includegraphics[scale=0.45]
    {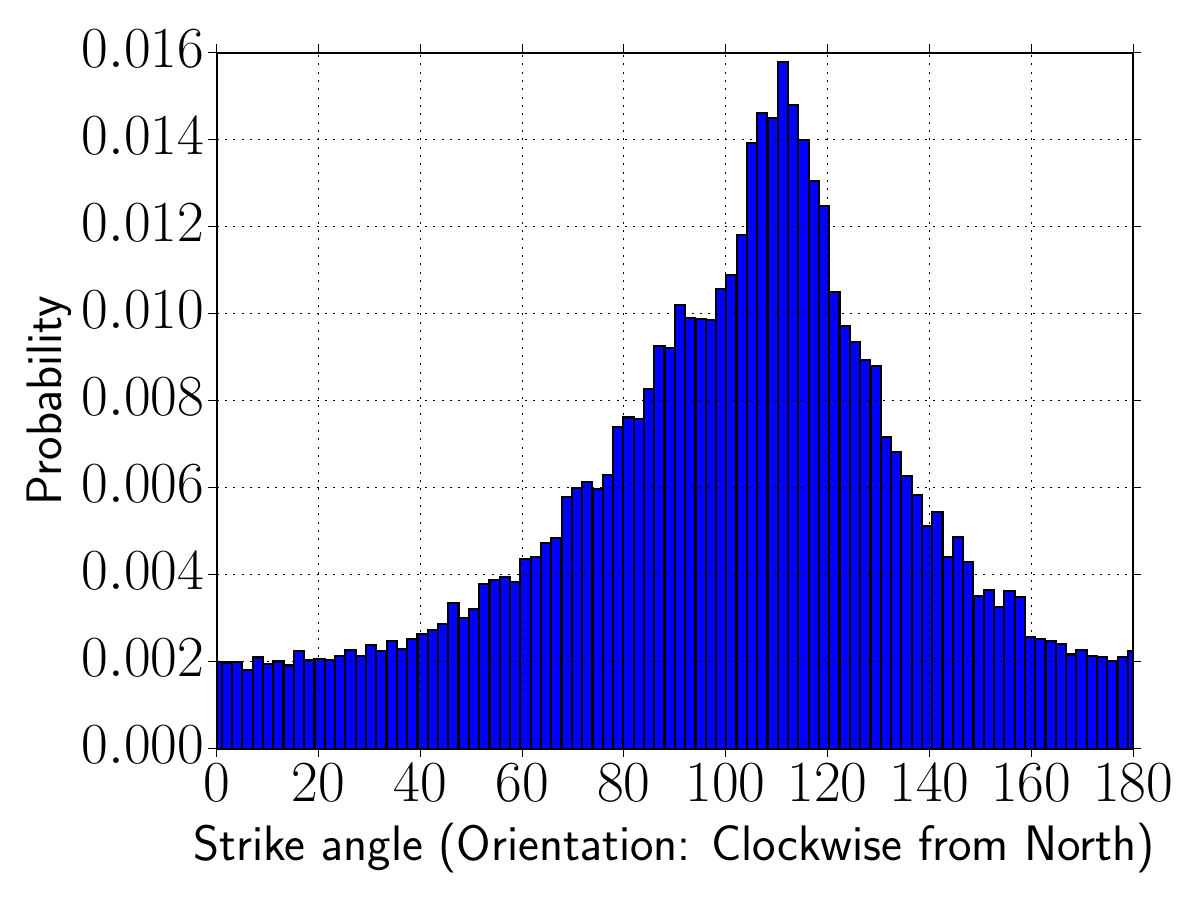}}
  \subfigure[Strike angle:~Cluster-3 (Inverted events)]
    {\includegraphics[scale=0.45]
    {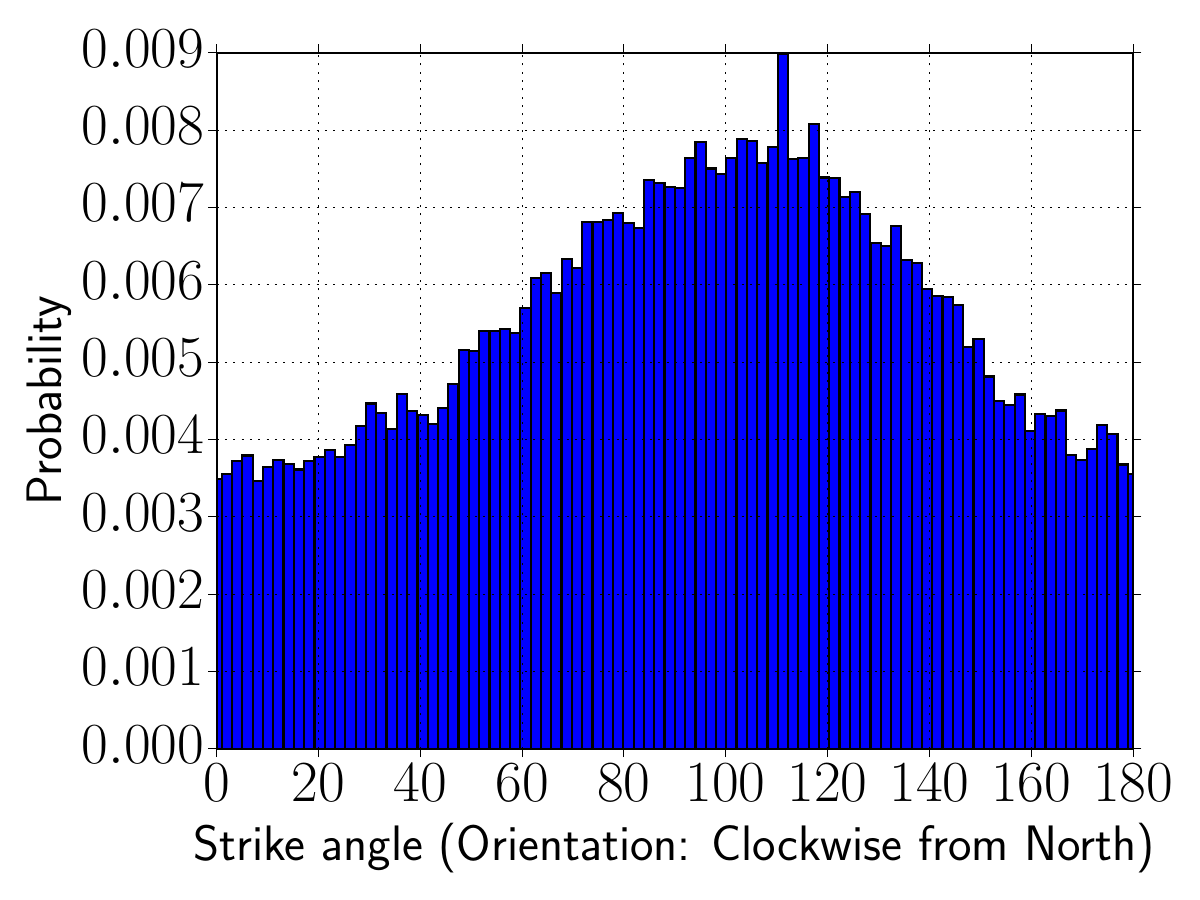}}
  \caption{\textsf{\textbf{Fault/fracture orientation (three 
    clusters):}}~Discrete probability distributions 
    for strike angle for inverted events based on clustering analysis.
  \label{Fig:JI_ClusteredData_Inv_1}}
\end{figure}

\begin{figure}
  \centering
  \subfigure[Dip angle:~Cluster-1 (Inverted events)]
    {\includegraphics[scale=0.45]
    {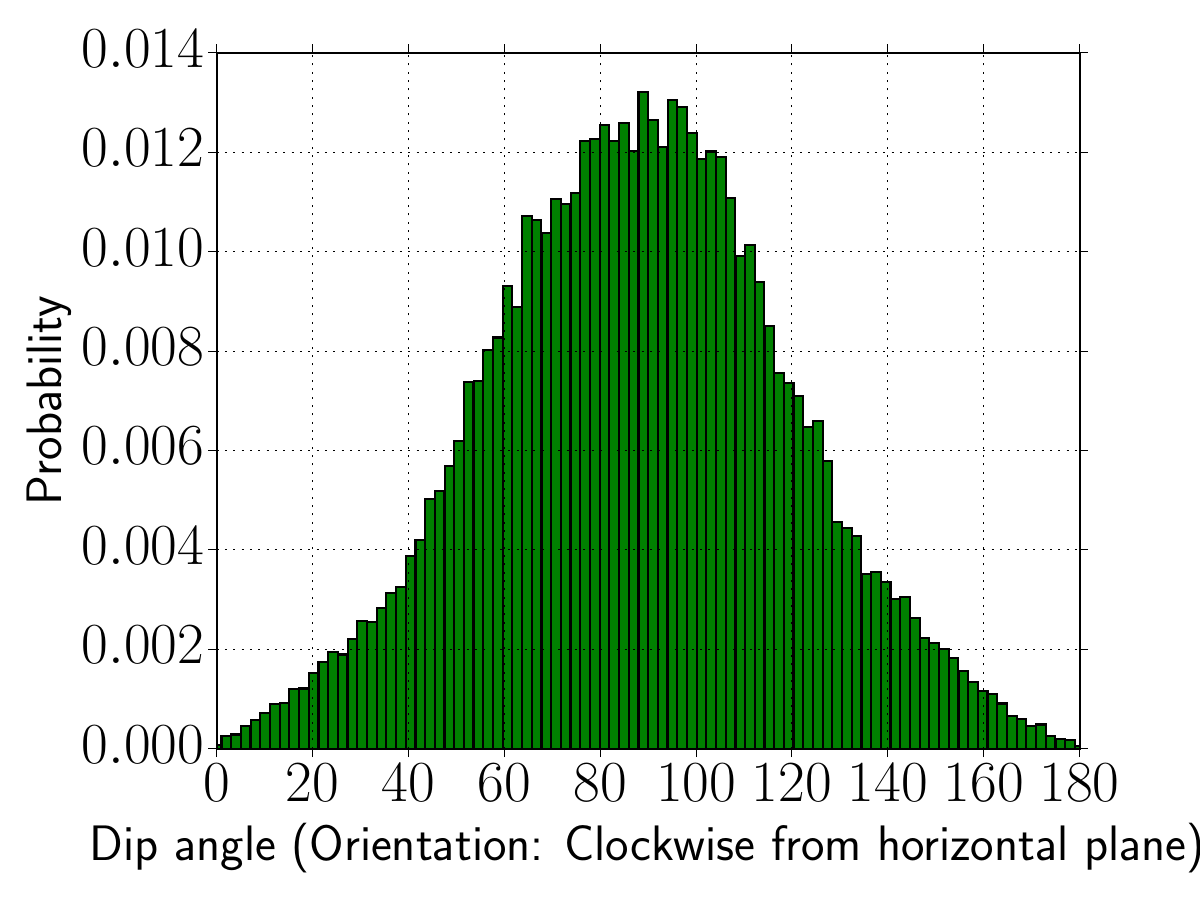}}
  \subfigure[Dip angle:~Cluster-2 (Inverted events)]
    {\includegraphics[scale=0.45]
    {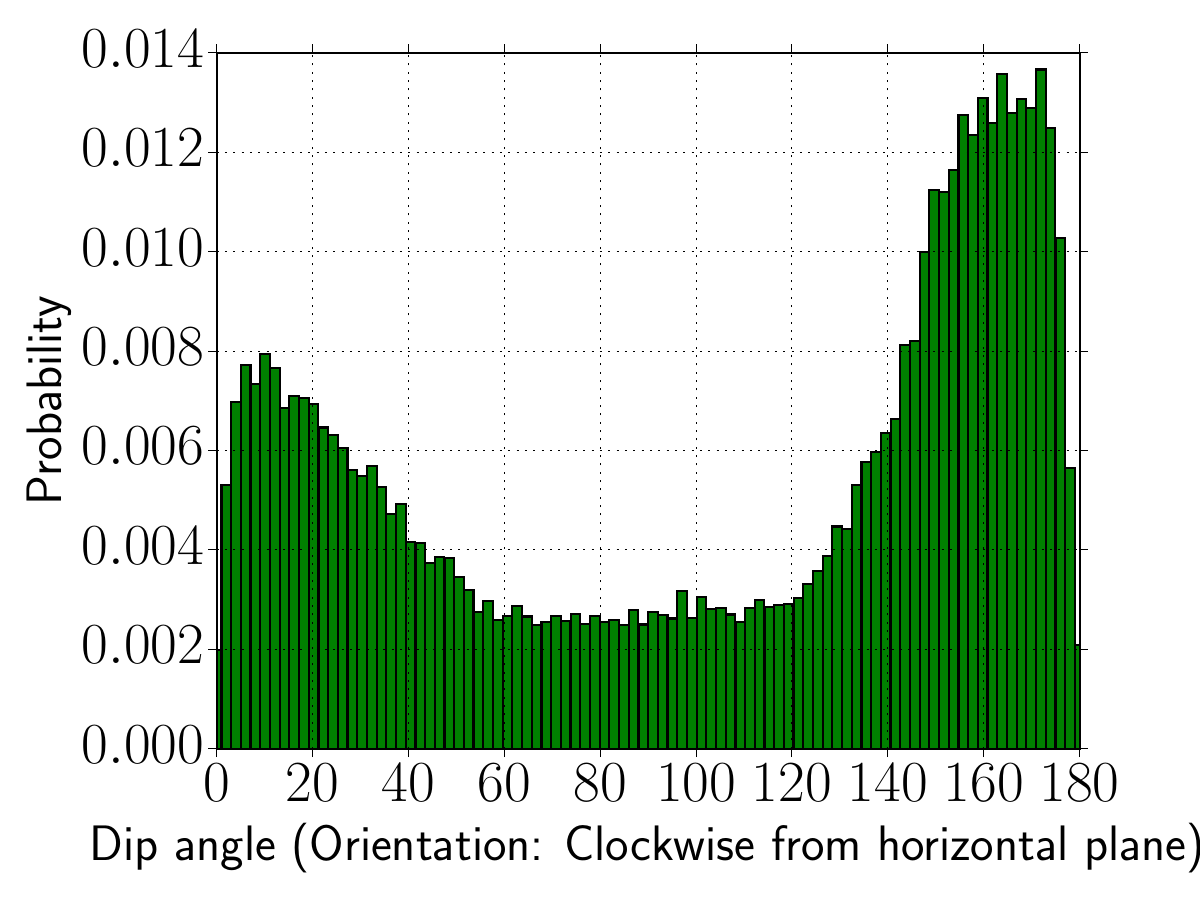}}
  \subfigure[Dip angle:~Cluster-3 (Inverted events)]
    {\includegraphics[scale=0.45]
    {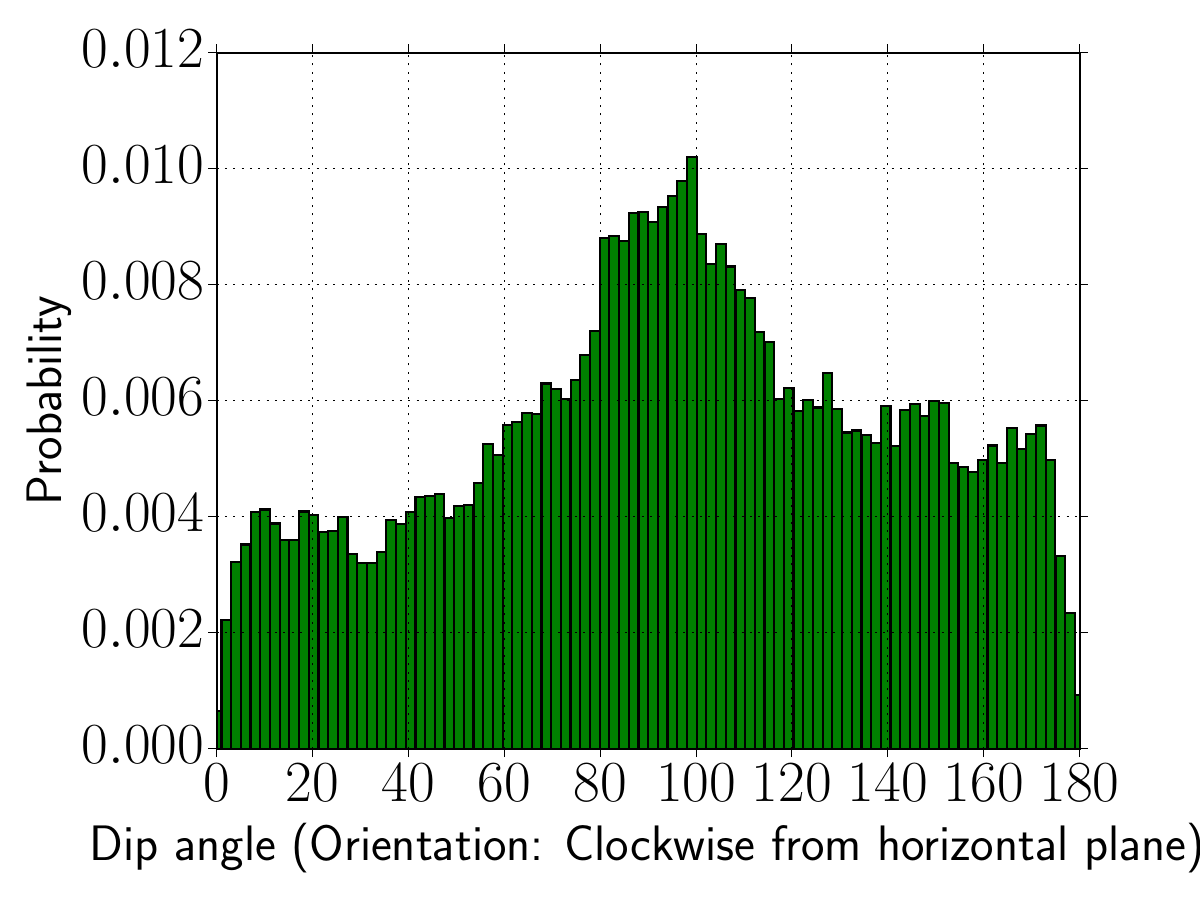}}
  \caption{\textsf{\textbf{Fault/fracture orientation (three 
    clusters):}}~Discrete probability distributions 
    for dip angle for inverted events based on clustering analysis.
  \label{Fig:JI_ClusteredData_Inv_2}}
\end{figure}

\begin{figure}
  \centering
  \subfigure[DFN mesh for three fractures]
    {\includegraphics[angle=-90, scale=0.55]
    {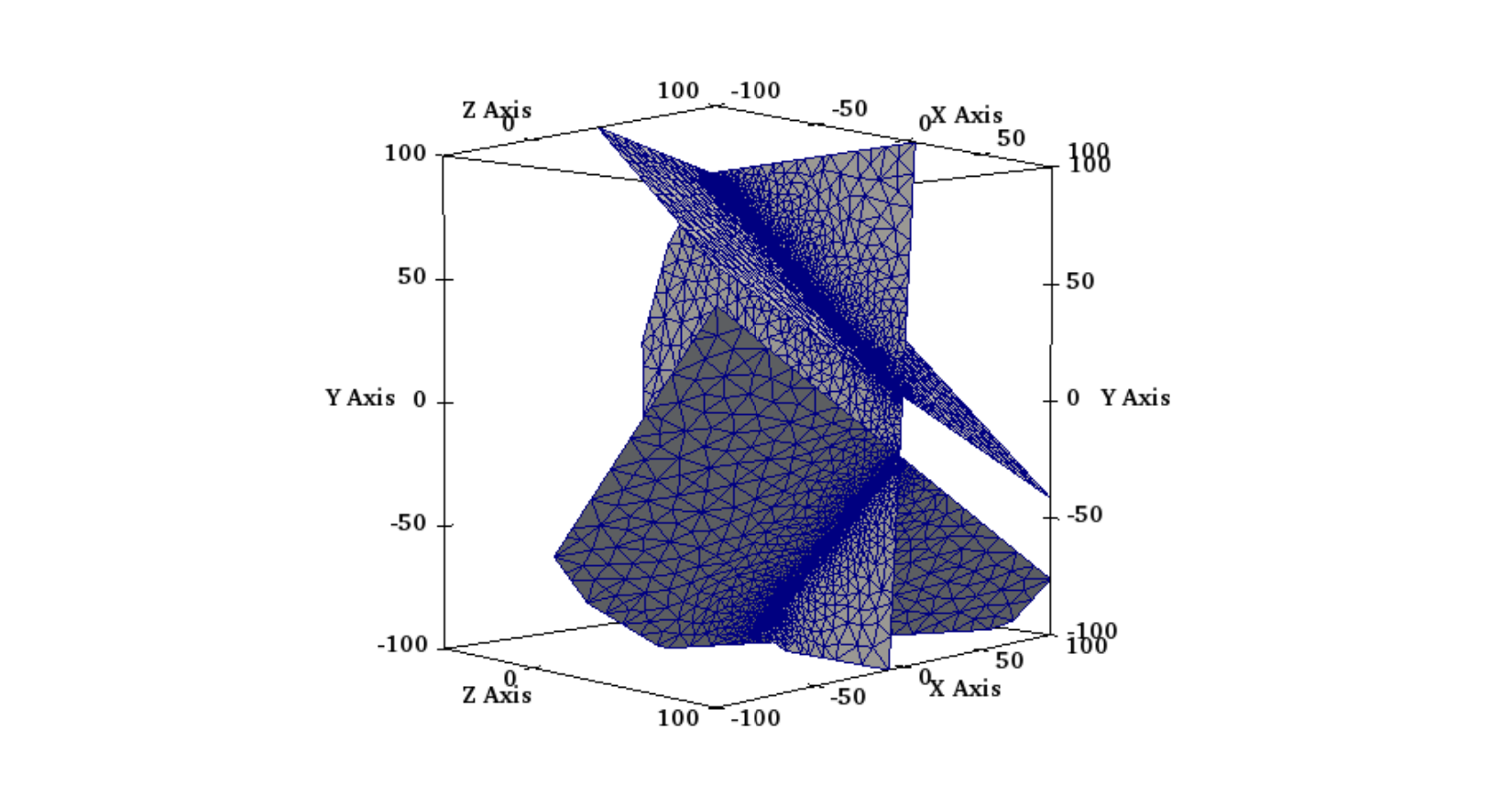}}
  \vspace{-0.1in}
  \subfigure[Liquid pressure profile for the corresponding 
    DFN (true solution)]
    {\includegraphics[angle=-90, scale=0.55]
    {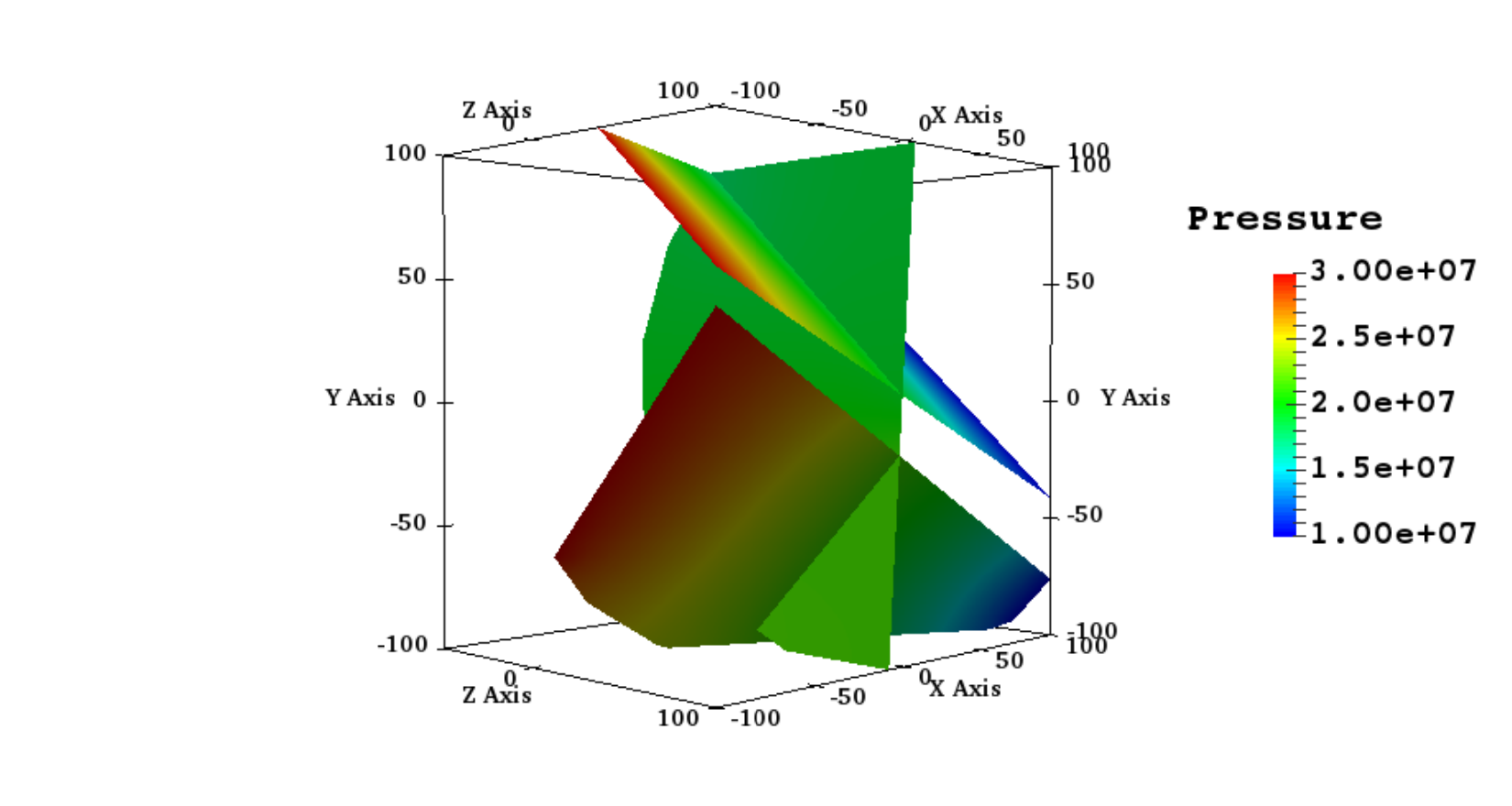}}
  \caption{\textsf{\textbf{Fault/fracture network with 
    three fractures (ground truth):}}~Figure (a) 
    shows the DFN mesh for three fractures constructed 
    using the unit normals provided in Section \ref{Sec:S3_JI_Results}.
    Figure (b) shows the profile of liquid pressure 
    on these three fractures. The fractures are drawn from 
    ellipses. The length of each fracture in the direction
    of minor axis is equal to 250 m while the aspect ratio (length 
    of major axis to minor axis) of these three fractures are 
    equal to 1.1, 1.2, and 1.25. The mean length for these 
    three fractures are 262.5 m, 275 m, and 281.25 m.
    The center of these elliptical fractures are located at 
    $(-5.543, -19.861, 98.218)$, $(0.577, 19.39, 91.1)$, and 
    $(9.42, 39.088, 53.548)$. Note that the domain of interest 
    is a cube of size $200 \times 200 \times 200 \; \mathrm{m}^3$ 
    and the reference datum for vertical depth is at 1500 m, which 
    is the top surface of the cube.
  \label{Fig:JI_DFN_Pressure_TrueSoln}}
\end{figure}

\begin{figure}
  \centering
  \subfigure[DFN constraint (Case \#1):~Constant aperture]
    {\includegraphics[angle=-90, clip, scale=0.6]
    {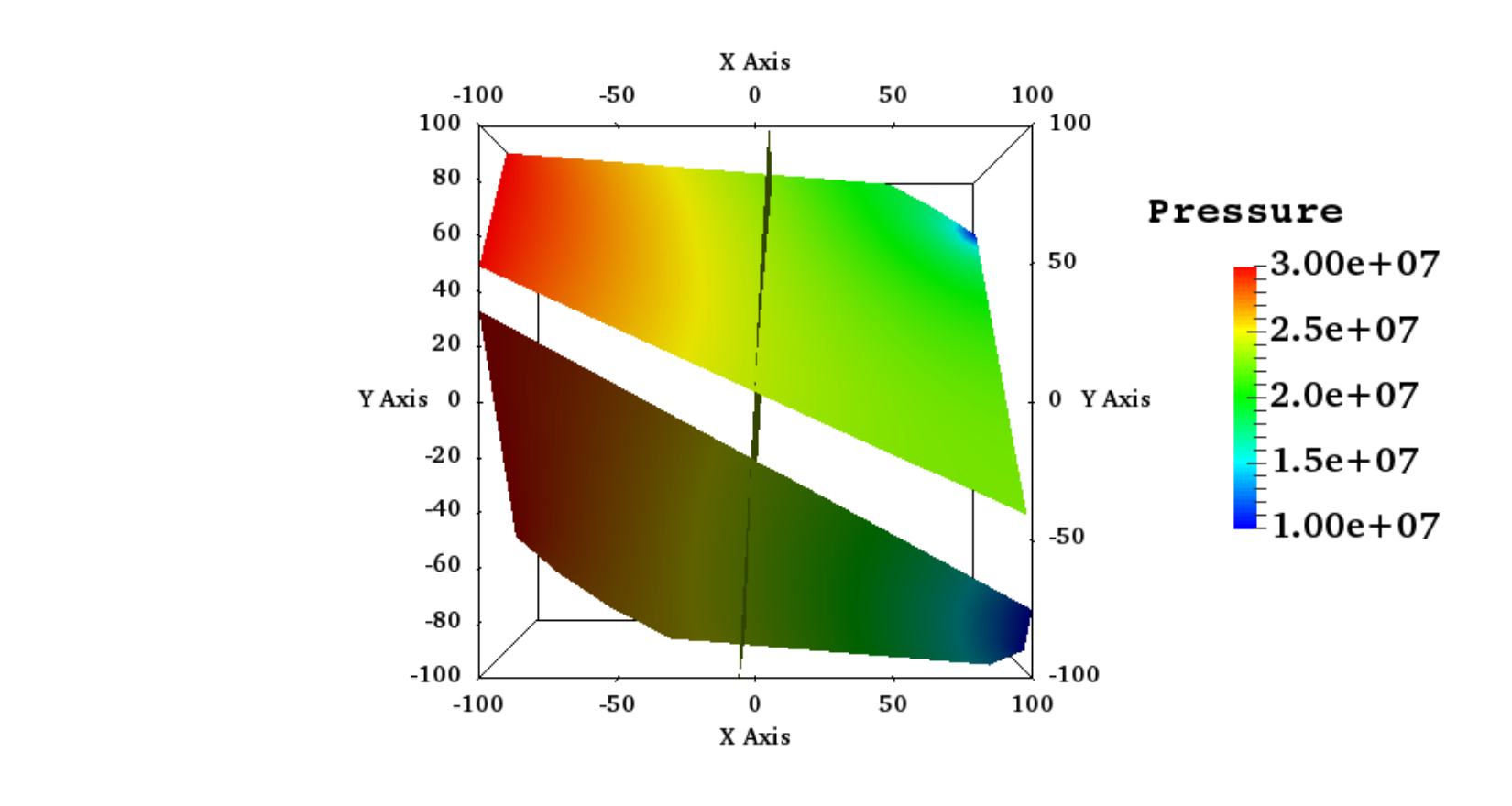}}
  \vspace{-0.15in}
  \subfigure[DFN constraint (Case \#2):~Log-normal aperture distribution]
    {\includegraphics[angle=-90, clip, scale=0.6]
    {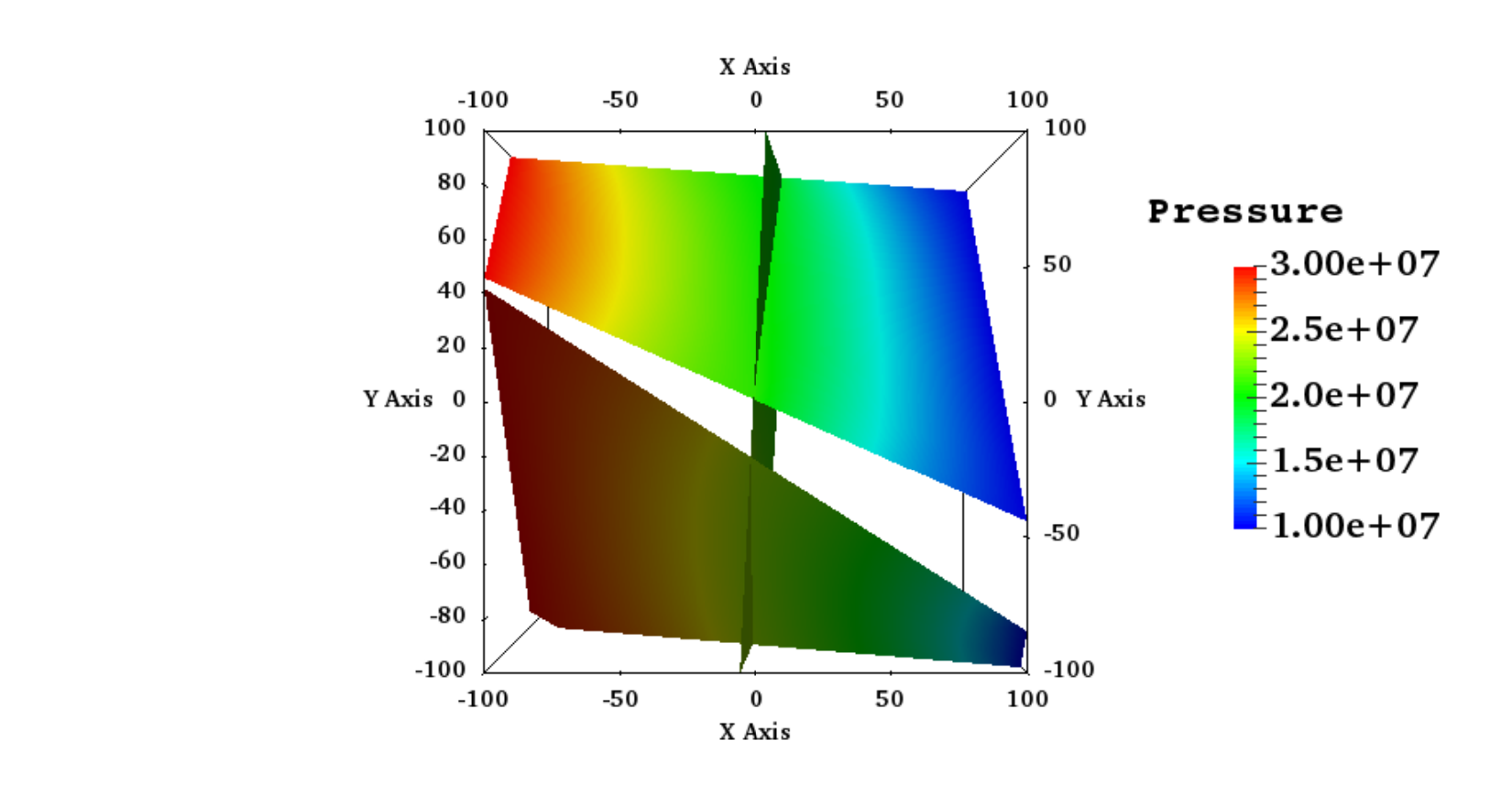}}
  \caption{\textsf{\textbf{Fault/fracture network with 
    three fractures (Case \#1 and Case \#2):}}~Figures 
    (a) and (b) show the profiles for 
    liquid pressure (for two different cases constant 
    aperture and log-normal aperture distribution), 
    which minimizes the misfit functional at the 
    observation points. Note that the liquid pressure 
    profiles are slightly different for these two cases.
    However, at the observation points these values are 
    close to each other. See Tables \ref{Tab:FracParams_1} 
    and \ref{Tab:FracParams_2} for fracture statistics.
  \label{Fig:JI_DFN_VariousCases1}}
\end{figure}

\begin{figure}
  \centering
  \subfigure[DFN constraint (Case \#3):~Fracture transmissivity and aperture correlation]
    {\includegraphics[angle=-90, clip, scale=0.6]
    {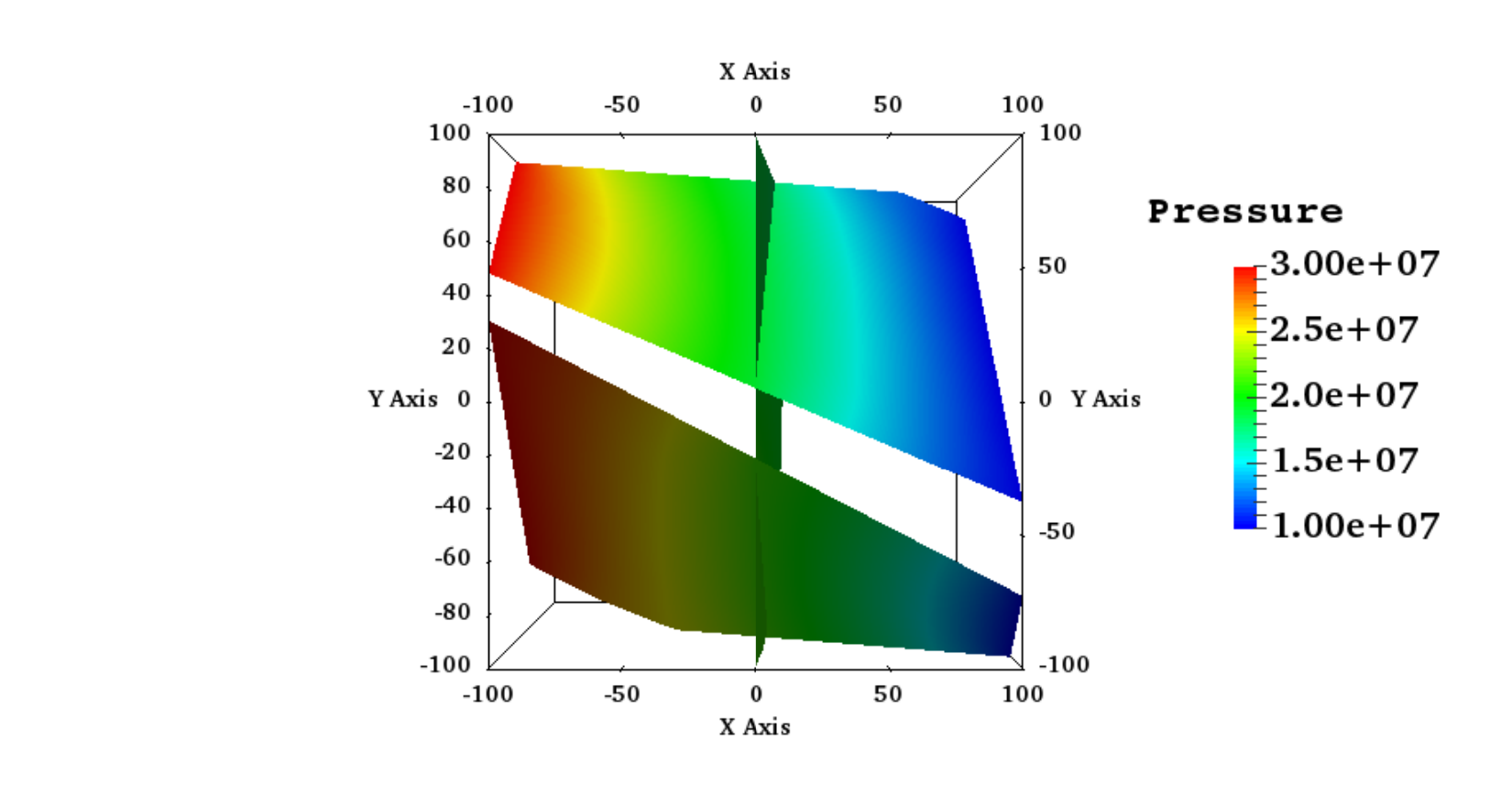}}
  \vspace{-0.15in}
  \subfigure[DFN constraint (Case \#4):~Fracture length and aperture correlation]
    {\includegraphics[angle=-90, clip, scale=0.6]
    {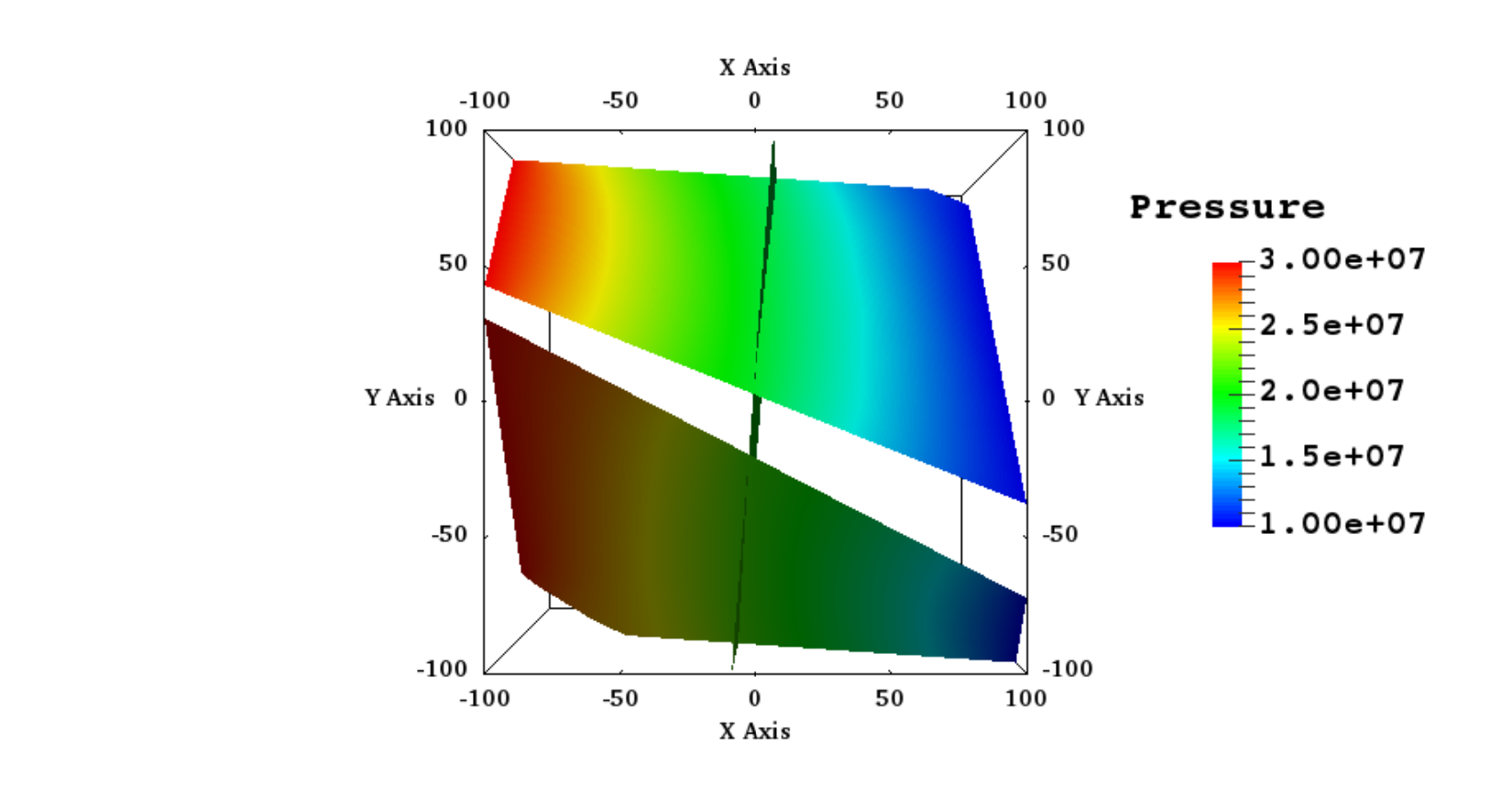}}
  \caption{\textsf{\textbf{Fault/fracture network 
    with three fractures (Case \#3 and Case \#4):}}~Figures 
    (a) and (b) show the liquid pressure 
    profiles based on fracture transmissivity-aperture 
    and fracture length-aperture correlations. See Tables \ref{Tab:FracParams_1} 
    and \ref{Tab:FracParams_2} for fracture statistics.
  \label{Fig:JI_DFN_VariousCases2}}
\end{figure}

Numerical results are shown in Figures \ref{Fig:JI_Ground_Truth_1}--\ref{Fig:JI_DFN_VariousCases2}.
From Figures \ref{Fig:JI_Ground_Truth_1}--\ref{Fig:JI_Ground_Truth_2}, 
the following can be inferred:~The 
strike and dip angles of true fracture planes are close to $(5^{\mathrm{o}}, 
90^{\mathrm{o}})$, $(117^{\mathrm{o}}, 60^{\mathrm{o}})$, and $(120^{\mathrm{o}}, 
120^{\mathrm{o}})$. From Figures \ref{Fig:JI_Entire_Cluster_2}--\ref{Fig:JI_ClusteredData_Inv_2}, 
which are constructed based on pure clustering analysis of inverted seismic events, 
a first set of upper and lower bounds on the strike and dip angles can be obtained. 
However, when used in conjunction with focal mechanisms. These bounds can be further 
constrained. To crystallize, from Figures 
\ref{Fig:JI_Entire_Cluster_1}--\ref{Fig:JI_ClusteredData_Inv_2}, the following conclusion 
can be drawn:~Based on focal mechanisms of inverted events, discrete probability 
distributions for strike and dip angles for entire cluster, and discrete probability 
distributions for strike and dip angles for each of the three different clusters, 
we have the following hard constraints on the fracture orientation:
\begin{description}
  \item[Fracture plane \#1] Strike and dip angles vary from 
    $0^{\mathrm{o}}-20^{\mathrm{o}}$ and $80^{\mathrm{o}}-100^{\mathrm{o}}$.
  \item[Fracture plane \#2] Strike and dip angles vary from 
    $100^{\mathrm{o}}-140^{\mathrm{o}}$ and $40^{\mathrm{o}}-80^{\mathrm{o}}$.
  \item[Fracture plane \#3] Strike and dip angles vary from 
    $100^{\mathrm{o}}-140^{\mathrm{o}}$ and $110^{\mathrm{o}}-140^{\mathrm{o}}$.
\end{description}

Latin hypercube samples are drawn from the above fracture angle ranges and the respective 
unit normals for fracture planes are obtained. These are provided as inputs for 
Algorithm \ref{Algo:DFN_Length} to estimate fracture length based on observation 
data at the observation points. Figure \ref{Fig:JI_DFN_Pressure_TrueSoln} provides 
the true solution for fluid pressure based on the true set of unit normals for 
elliptical fractures. Numerical simulations are performed for the four different 
cases based on the constructed LHS. Figures \ref{Fig:JI_DFN_VariousCases1} 
and \ref{Fig:JI_DFN_VariousCases2} provide the contour profiles for fluid pressure 
for which the misfit functional value is in the order $10^{-4}$. For Case \#3 and 
Case \#4, the parameter $\alpha$ is equal to 0.75 and 0.804.

\begin{table}
  \centering
	\caption{Fracture aperture and the relative 
	  error for four different cases.
	\label{Tab:FracParams_1}}
	\begin{tabular}{|c|c|c|c|c|c|c|} \hline
	  \multirow{2}{*}{{\scriptsize Case \#}} &  
	  \multicolumn{3}{|c|}{{\scriptsize Aperture [m]}} &
	  \multicolumn{3}{|c|}{{\scriptsize $\mathrm{Relative \; error} 
	  = \frac{\mathrm{Aperture \; value}}{\mathrm{Ground \; 
	  truth \; value}} - 1$}} \\
	  \cline{2-7}
	  & {\scriptsize Frac-1} & {\scriptsize Frac-2} & {\scriptsize 
	  Frac-3} & {\scriptsize Frac-1} & {\scriptsize Frac-2} & {\scriptsize 
	  Frac-3} \\ \hline
	  {\scriptsize True value} & {\scriptsize $10^{-5}$} & 
	  {\scriptsize $10^{-5}$} 
	  & {\scriptsize $10^{-5}$} & 
	  {\scriptsize NA} & 
	  {\scriptsize NA} & 
	  {\scriptsize NA} \\
	  {\scriptsize 1} & {\scriptsize $2.2569 \times 10^{-5}$} 
	  & {\scriptsize $2.2569 \times 10^{-5}$} 
	  & {\scriptsize $2.2569 \times 10^{-5}$} 
	  & {\scriptsize $1.26$}
	  & {\scriptsize $1.26$}
	  & {\scriptsize $1.26$} \\
	  {\scriptsize 2} & {\scriptsize $5.3222 \times 10^{-6}$}
	  & {\scriptsize $4.6028 \times 10^{-6}$}
	  & {\scriptsize $3.585 \times 10^{-6}$} 
	  & {\scriptsize $-0.47$}
	  & {\scriptsize $-0.54$}
	  & {\scriptsize $-0.64$} \\
	  {\scriptsize 3} & {\scriptsize $4.2092 \times 10^{-5}$} 
	  & {\scriptsize $4.2256 \times 10^{-5}$} 
	  & {\scriptsize $4.2454 \times 10^{-5}$} 
	  & {\scriptsize $3.21$} 
	  & {\scriptsize $3.23$} 
	  & {\scriptsize $3.25$} \\
	  {\scriptsize 4} & {\scriptsize $2.596 \times 10^{-3}$} 
	  & {\scriptsize $2.722 \times 10^{-3}$} 
	  & {\scriptsize $2.832 \times 10^{-3}$}
	  & {\scriptsize $2.56 \times 10^{2}$} 
	  & {\scriptsize $2.72 \times 10^{2}$} 
	  & {\scriptsize $2.83 \times 10^{2}$} \\
	  \hline
	\end{tabular}
\end{table}

\begin{table}
  \centering
	\caption{Fracture permeability and relative error 
	for four different cases.
	\label{Tab:FracParams_2}}
	\begin{tabular}{|c|c|c|c|c|c|c|} \hline
      \multirow{2}{*}{{\scriptsize Case \#}} &  
	  \multicolumn{3}{|c|}{{\scriptsize Permeability $[\mathrm{m}^2]$}} &
	  \multicolumn{3}{|c|}{{\scriptsize $\mathrm{Relative \; error} 
	  = \frac{\mathrm{Permeability \; value}}{\mathrm{Ground \; 
	  truth \; value}} - 1$}} \\
	  \cline{2-7}
	  & {\scriptsize Frac-1} & {\scriptsize Frac-2} & {\scriptsize 
	  Frac-3} & {\scriptsize Frac-1} & {\scriptsize Frac-2} & {\scriptsize 
	  Frac-3} \\ \hline
	  {\scriptsize True value} & {\scriptsize $10^{-12}$} & 
	  {\scriptsize $10^{-12}$} 
	  & {\scriptsize $10^{-12}$} & 
	  {\scriptsize NA} & 
	  {\scriptsize NA} & 
	  {\scriptsize NA} \\
	  {\scriptsize 1} & {\scriptsize $1.716 \times 10^{-13}$} 
	  & {\scriptsize $1.716 \times 10^{-13}$} 
	  & {\scriptsize $1.716 \times 10^{-13}$} 
	  & {\scriptsize $-0.83$}
	  & {\scriptsize $-0.83$}
	  & {\scriptsize $-0.83$} \\
	  {\scriptsize 2} & {\scriptsize $2.361 \times 10^{-12}$}
	  & {\scriptsize $1.765 \times 10^{-12}$}
	  & {\scriptsize $1.071 \times 10^{-12}$} 
	  & {\scriptsize $1.36$}
	  & {\scriptsize $0.77$}
	  & {\scriptsize $0.07$} \\
	  {\scriptsize 3} & {\scriptsize $1.476 \times 10^{-10}$}
	  & {\scriptsize $1.488 \times 10^{-10}$}
	  & {\scriptsize $1.502 \times 10^{-10}$}
	  & {\scriptsize $1.47 \times 10^{2}$}
	  & {\scriptsize $1.48 \times 10^{2}$}
	  & {\scriptsize $1.49 \times 10^{2}$} \\
	  {\scriptsize 4} & {\scriptsize $5.617 \times 10^{-7}$}
	  & {\scriptsize $6.178 \times 10^{-7}$}
	  & {\scriptsize $6.685 \times 10^{-7}$} 
	  & {\scriptsize $5.62 \times 10^{5}$}
	  & {\scriptsize $6.18 \times 10^{5}$}
	  & {\scriptsize $6.69 \times 10^{5}$} \\
	  \hline
	\end{tabular}
\end{table}

\begin{table}
  \centering
	\caption{Model observational pressures and relative 
	error for four different cases.
	\label{Tab:FracParams_3}}
	\begin{tabular}{|c|c|c|c|c|c|c|} \hline
	  \multirow{2}{*}{{\scriptsize Case \#}} &  
	  \multicolumn{3}{|c|}{{\scriptsize Pressure [MPa]}} &
	  \multicolumn{3}{|c|}{{\scriptsize $\mathrm{Relative \; error} 
	  = \frac{\mathrm{Pressure \; value}}{\mathrm{Ground \; 
	  truth \; value}} - 1$}} \\
	  \cline{2-7}
      & {\scriptsize Frac-1} & {\scriptsize Frac-2} & {\scriptsize 
	  Frac-3} & {\scriptsize Frac-1} & {\scriptsize Frac-2} & {\scriptsize 
	  Frac-3} \\ \hline
	  {\scriptsize True value} & {\scriptsize 21.86}
	  & {\scriptsize 19.08}
	  & {\scriptsize 18.33} &
	  {\scriptsize NA} & 
	  {\scriptsize NA} & 
	  {\scriptsize NA}\\
	  {\scriptsize 1} & {\scriptsize 21.74} 
	  & {\scriptsize 19.04}
	  & {\scriptsize 18.29} 
	  & {\scriptsize -0.0054} 
	  & {\scriptsize -0.0021}
	  & {\scriptsize -0.0022} \\
	  {\scriptsize 2} & {\scriptsize 21.49} 
	  & {\scriptsize 19.96}
	  & {\scriptsize 18.58} 
	  & {\scriptsize -0.017} 
	  & {\scriptsize 0.046}
	  & {\scriptsize 0.014} \\
	  {\scriptsize 3} & {\scriptsize 22.09} 
	  & {\scriptsize 19.02}
	  & {\scriptsize 18.19}
	  & {\scriptsize 0.011} 
	  & {\scriptsize -0.003}
	  & {\scriptsize -0.007} \\
	  {\scriptsize 4} & {\scriptsize 21.52} 
	  & {\scriptsize 18.57}
	  & {\scriptsize 17.96}
	  & {\scriptsize -0.015} 
	  & {\scriptsize -0.027}
	  & {\scriptsize -0.008} \\
	  \hline
	\end{tabular}
\end{table}

\begin{table}
  \centering
	\caption{Axes length and relative error 
	  for four different cases.
	\label{Tab:FracParams_4}}
	\begin{tabular}{|c|c|c|c|c|c|c|c|c|} \hline
	  \multirow{2}{*}{{\scriptsize Case \#}}{} &  
	  \multicolumn{4}{|c|}{{\scriptsize Axis length [m]}} &
	  \multicolumn{4}{|c|}{{\scriptsize $\mathrm{Relative \; error} 
	  = \frac{\mathrm{Axis length \; value}}{\mathrm{Ground \; 
	  truth \; value}} - 1$}} \\
	  \cline{2-9}
	  & {\scriptsize Minor (all)} & {\scriptsize 
	  Frac-1 (Major)} & {\scriptsize Frac-2 (Major)} & {\scriptsize 
	  Frac-3 (Major)} & {\scriptsize Minor (all)} & {\scriptsize 
	  Frac-1} & {\scriptsize Frac-2} & {\scriptsize 
	  Frac-3} \\ \hline
	  {\scriptsize True value} & {\scriptsize 250} 
	  & {\scriptsize 275} 
	  & {\scriptsize 300}
	  & {\scriptsize 312.5}
	  & {\scriptsize NA} 
	  & {\scriptsize NA} 
	  & {\scriptsize NA}
	  & {\scriptsize NA} \\
	  {\scriptsize 1} & {\scriptsize 245.14}
	  & {\scriptsize 255.88} 
	  & {\scriptsize 255.44} 
	  & {\scriptsize 263.34}
	  & {\scriptsize -0.019}
	  & {\scriptsize -0.069} 
	  & {\scriptsize -0.149} 
	  & {\scriptsize -0.157} \\
	  {\scriptsize 2} & {\scriptsize 280.12}
	  & {\scriptsize 335.62} 
	  & {\scriptsize 301.28}
	  & {\scriptsize 338.1}
	  & {\scriptsize 0.121}
	  & {\scriptsize 0.221} 
	  & {\scriptsize 0.004}
	  & {\scriptsize 0.082} \\
	  {\scriptsize 3} & {\scriptsize 245.98} 
	  & {\scriptsize 296.24}
	  & {\scriptsize 300.48}
	  & {\scriptsize 306.56} 
	  & {\scriptsize -0.016} 
	  & {\scriptsize 0.077}
	  & {\scriptsize 0.002}
	  & {\scriptsize -0.019}\\
	  {\scriptsize 4} & {\scriptsize 250.64}
	  & {\scriptsize 272.64} 
	  & {\scriptsize 289.3}
	  & {\scriptsize 303.84} 
	  & {\scriptsize 0.002}
	  & {\scriptsize -0.008} 
	  & {\scriptsize -0.035}
	  & {\scriptsize -0.027}\\
	  \hline
	\end{tabular}
\end{table}

A summary of the above parameters are shown in Tables 
\ref{Tab:FracParams_1}, \ref{Tab:FracParams_2}, and 
\ref{Tab:FracParams_4}. The following can be inferred 
based on Figures \ref{Fig:JI_Ground_Truth_1}--\ref{Fig:JI_DFN_VariousCases2} 
and Tables \ref{Tab:FracParams_1}--\ref{Tab:FracParams_4}:
\begin{description}
  \item [Fracture orientation]~Pure clustering analysis (statistical 
    approach) without the information from focal mechanisms (physics-based 
    approach) may not always be accurate. But it is the combination of 
    clustering analysis of seismic events and focal mechanisms that 
    can provide reasonably accurate information, thereby ensuring 
    hard constraints on the dominant fracture planes orientation. 
    Details on how we arrived at the fracture angles 
    and number of fracture planes based on a combination of cluster 
    analysis and focal mechanisms are given below:
    \begin{itemize}
      \item From elbow method \cite{2013_Kodinariya_Makwana_IJ_v1_p90_p95,
        2005_Tibshirani_Walther_v14_p511_p528,2001_Tibshirani_etal_v63_p411_p423}, 
        which is pure clustering analysis, it is clear that we can have 
        three or four or five possible fracture planes. 
      \item From strike angle distribution obtained from the entire 
        cluster analysis (Figures \ref{Fig:JI_Entire_Cluster_2}--\ref{Fig:JI_Entire_Cluster_3}) 
        of inverted seismic events, we observe that we have three 
        possible clusters. However, dip angle distribution obtained 
        from clustering analysis doesn't provide much information.
      \item Strike angle distribution obtained from focal mechanism
        shows two dominant fracture planes. Dip angle distribution 
        shows three dominant fracture planes.
      \item Based on the elbow method and focal mechanisms, we can 
        conclude that there are three dominant fracture planes.  
    \end{itemize}
   \item [Fracture aperture and permeability]~From Table \ref{Tab:FracParams_1}, it 
     can inferred that Case \#2 has aperture values close to ground 
     truth while Case \#4 performs poorly. From Table \ref{Tab:FracParams_2}, 
     Case \#1 and Case \#2 have permeability values close to ground 
     truth while Case \#3 and Case \#4 deviate considerably. This is 
     due to the nonlinear relationships between fracture aperture
     and fracture length assumed in these two cases.
  \item [Model pressure values]~From Table \ref{Tab:FracParams_3}, 
    for all the cases the model pressure values are close to each 
    other. This is expected as it is a consequence of Equation 
    \ref{Eqn:Misfit_Functional}. From Figures \ref{Fig:JI_DFN_VariousCases1} 
    and \ref{Fig:JI_DFN_VariousCases2}, qualitatively, the pressure 
    profiles for Case \#2, \#3, and \#4 are similar to ground truth.
  \item [Fracture lengths]~From Table \ref{Tab:FracParams_4}, the 
    Case \#4 has least absolute relative error for minor axis while 
    Case \#2 has highest absolute relative error. For major axis, 
    for fracture-1, Case \#4 has least error while Case \#2 has the 
    highest error. For fracture-2 and fracture-3, Case \#3 has the 
    least error while Case \#1 has the highest absolute error.
    In all the cases, the absolute relative error for both major 
    and minor axis are in the order of $10^{-2}$. 
\end{description}

%

\section{CONCLUDING REMARKS}
\label{Sec:S4_JI_Conclusions}
In this paper, we have presented a sequential inversion framework to 
constrain subsurface fracture network represented as a
DFN using geophysical and flow data sets. First, we described the forward 
models to obtain focal mechanisms and model observation data for 
flow variables. Second, based on these models we have constructed 
a novel sequential inversion methodology for various cases to constrain 
dominant fracture orientation and fracture length. Third, utilizing 
this framework, we have presented a synthetic numerical example to 
demonstrate various aspects of the proposed algorithms. From this 
example, it is apparent that physics-informed clustering analysis,
which is a combination of clustering analysis (pure statistical 
approach) and focal mechanisms (physics-based approach) can provide 
accurate bounds on the dominant fracture plane orientations. Finally, 
we discussed various cases to constrain fracture length. From these 
case studies, it is clear that the model flow data variables are 
highly dependent on various fracture parameters (which inturn depend 
on the fracture length in a highly nonlinear fashion). In all these 
case studies, even though the model observation data is close to 
the prescribed observation data, there is a wide difference in 
different fracture parameters. Furthermore, prior analysis on these 
fracture parameters (such as aperture, transmissivity, and permeability) 
is needed to better constrain the fracture length.

It is seen from the example problem 
that the resulting primary variables (such as pressure and flow 
rate), obtained by solving the governing equations for the flow 
problem, are very sensitive to the fracture parameters (aperture,
 transmissivity and its coefficients, and permeability), 
fracture length, and fracture orientation. Moreover, fracture 
parameters such as fracture permeability have a nonlinear dependence 
on the fracture length. This has serious consequences on constraining 
subsurface DFNs as pressure and flow rate depend on the fracture 
permeability. Hence, it is a challenge to obtain sharp constraints 
on the fracture statistics. 

The proposed framework and algorithms are general and non-intrusive. 
They leverage on existing parallel fracture flow simulators 
to construct bounds on the fracture length and orientation. Moreover, 
extending the sequential inversion framework proposed 
in this paper to include various other nearly-orthogonal/complementary 
datasets (such as chemical, mechanical, and temperature) is straight-forward 
and requires minimal effort.    

Subsequently, in our future works, we will perform detailed sensitivity 
analysis on these fracture parameters to impose hard/sharp bounds on the 
fracture length, aperture, transmissivity, and permeability.
As the number of fractures increases, more sophisticated methods such as
$X$-means clustering, information-theoretic approach, silhouette method, 
and kernel matrix analysis can be used to determine the number of cluster/fractures 
\cite{2013_Kodinariya_Makwana_IJ_v1_p90_p95} rather than relying on traditional 
elbow method. 
%

\section*{APPENDIX}
A brief description of the geophysical, subsurface 
flow, and statistical terminology used in this paper, is given below:
\begin{description}
  \item[Discrete Fracture Network (DFN) modeling]~The DFN approach 
    is a modeling methodology that seeks to describe the rock mass 
    fracture system in statistical ways by building a series 
    of discrete fracture objects represented as two-dimensional 
    planes in three-dimensional space. The statistics of these
    networks are based on field observations of fracture properties 
    such as size, orientation and intensity. 
  \item[Outcrop]~An outcrop is a visible exposure of bedrock 
    or superficial deposits on the surface of the earth.
  \item[Focal mechanism]~Seismologists refer to the direction 
    of slip in an earthquake and the orientation of the fault 
    on which it occurs as the focal mechanism. They use the 
    information from seismograms to calculate the focal mechanism.
  \item[Microseismics]~In seismology, a microseism is defined 
    as a faint earth tremor caused by natural phenomena. Microseisms 
    are very well detected and measured by means of a broad-band 
    seismograph, and can be recorded anywhere on Earth.
  \item[Shear wave]~In seismology, shear waves are one of the 
    two main types of elastic body waves, because they move 
    through the bulk of the object/body under consideration, 
    unlike surface waves.
  \item[Permeability]~Permeability in fluid mechanics and the 
    earth sciences is a measure of the ability of a porous material 
    to allow fluids to pass through it.
  \item[Fracture aperture]~Fracture aperture is the  
    width of a fracture opening. In quantifying flow through fractures, 
    apertures can be used to calculate permeability by assuming 
    equivalency with flow between two parallel plates.
  \item[Compensated Linear Vector Dipole]~The pattern of energy 
    radiation of an earthquake is represented by the moment tensor 
    solution, which is graphically represented by beachball diagrams. 
    An explosive or implosive mechanism produces an isotropic seismic 
    source. Slip on a planar fault surface results in what is known 
    as a double-couple source. Uniform outward motion in a single plane 
    due to normal shortening is known as a compensated linear vector 
    dipole source.
  \item[Geophones]~A geophone is a device that converts ground movement 
    (velocity) into voltage, which may be recorded at a recording station. 
    The deviation of this measured voltage from the base line is called 
    the seismic response and is analyzed for structure of the earth.
  \item[Seismic moment]~Seismic moment is a quantity used by earthquake 
    seismologists to measure the size of an earthquake. For modern earthquakes, 
    moment is usually estimated from ground motion recordings of earthquakes 
    known as seismograms. For earthquakes that occurred in times before modern 
    instruments were available, moment may be estimated from geologic estimates 
    of the size of the fault rupture and the displacement.
  \item[Green's function]~Green's function visualizes the effect of source 
    concentrated at a point on different points of the domain. In mathematics, 
    a Green's function is the impulse response of an inhomogeneous differential 
    equation defined on a domain, with specified initial conditions or boundary 
    conditions. Combining Green's function with focal mechanism can represent seismic 
    waveforms recorded for a given earthquake. 
  \item[$k$-means clustering]~Clustering is a process of partitioning a set 
    of data (or objects) into a set of meaningful sub-classes, called clusters. 
    This helps user to understand the natural grouping or structure in a dataset. 
    Clustering is used either as a stand-alone tool to get insight into data distribution 
    or as a pre-processing step for other algorithms. In particular, $k$-means 
    clustering aims to partition $n$ observations into $k$ clusters in which each 
    observation belongs to the cluster with the nearest mean, serving as a prototype 
    of the cluster \cite{2002_Kanungo_etal_IEEE_TPAMI_v24_p881_p892,
    2004_Setodji_etal_Techno_v46_p421_p429,2010_Jain_PRL_v31_p651_p666}.
  \item[Latin Hypercube Sampling]~Latin Hypercube Sampling (LHS) is a statistical 
    method for generating a near-random sample of parameter values from a multidimensional 
    distribution. The sampling method is often used to construct computer experiments 
    \cite{1979_McKay_etal_Techno_v21_p239_p245,1979_McKay_Techno_v21_p475_p479}. 
  \item[Sequential geophysical and flow inversion]~By this term, we mean that the 
    geophysical data is inverted first. This inversion process results in fracture 
    orientations. Then, we invert for flow data to obtain fracture lengths.
\end{description}

\section*{ACKNOWLEDGMENTS}
The authors thank U.S. Department of Energy (DOE) SubTER initiative. 
MKM and SK acknowledge the support of the LANL LDRD Early Career 
Project 20150693ECR. MKM thanks Hari Viswanathan, Gilles 
Bussod, and Vamshi Chillara for many useful discussions. 
MKM also thanks Daniel O'Malley for the help related to \textsf{MADS} 
software. TC thanks Yu Chen for providing the focal mechanism inversion 
code.

\bibliographystyle{unsrt}
\bibliography{Master_References/Master_References,Master_References/Books}
%


\end{document}